\documentstyle[pre,aps]{revtex} 
\begin{document}
\newcommand{\bea}{\begin{eqnarray}}
\newcommand{\eea}{\end{eqnarray}}

\title{The $U_q\bigl(\widehat{sl}(2\vert1)\bigr)_1$-module
$V(\Lambda_2)$ and a Corner Transfer Matrix at $q=0$}
\author{R.M. Gade\footnote{e-mail:renate.gade@t-online.de}
\\Lerchenfeldstra\ss e 12, 80538 Munich, Germany}
\maketitle

\begin{abstract}
The north-west corner transfer matrix of an inhomogeneous
integrable vertex model constructed from
the vector representation of $U_q\bigl(sl(2\vert1)\bigr)$ and its dual
is investigated. In the limit $q\to0$, the spectrum can be obtained. Based
on an analysis of the half-infinite tensor products related to all
CTM-eigenvalues $\geq-4$, it is argued that the eigenvectors of the corner
transfer matrix are in one-to-one correspondence with the weight state
of the $U_q\bigl(\widehat{sl}(2\vert1)\bigr)$-module $V(\Lambda_2)$ at
level one. This is supported by a comparison of the complete set of
eigenvectors with a nondegenerate triple of eigenvalues of the
CTM-Hamiltonian and the generators of the Cartan-subalgebra of
$U_q\bigl(sl(2\vert1)\bigr)$ to the weight states of $V(\Lambda_2)$
with multiplicity one.
\end{abstract}

\vskip 0.5cm
Keywords: Integrable models, quantum affine superalgebras

\section{Introduction}

A variety of models with an underlying $gl(2\vert1)$-symmetry and
their anisotropic generalizations with
$U_q\bigl(gl(2\vert1)\bigr)$-invariance (for references see \cite{goe}
and \cite{sed}) has been receiving attention following the introduction
of the Perk-Schultz model \cite{ps} (or supersymmetric $t-J$ model
\cite{s}). The model specified in
\cite{ps} is based on the R-matrix associated to the three-dimensional
vector representation of $gl(2\vert1)$. Later, models constructed from
other $gl(2\vert1)$- or $U_q\bigl(gl(2\vert1)$-representations have
been investigated, among these
the generalized supersymmetric Hubbard model related to the
four-dimensional representation (\cite{ks} and references therein)
or a family of "doped Heisenberg chains" involving
atypical representations \cite{fr}. 

More recently, models incorporating
various types of inhomogeneities have been considered.
These include impurity $t-J$ models with the vector representation
at one site (or few sites) of the quantum space substituted
by a four-dimensional representation \cite{bef},\cite{foe} and periodic
inhomogeneities. The latter may be implemented as a staggered disposition
of the spectral parameter \cite{frr}, \cite{zv}
or both the spectral and the anisotropy parameter {\cite{sed}.
Another possibility consists in combining several representations
in a periodic sequence. In \cite{LF}, \cite{ar} the analysis of a model
composed of the vector representation of $sl(2\vert1)$ and its dual
is addressed to by means of the algebraic Bethe ansatz. Using the
functional Bethe ansatz, the thermodynamic behavior of  
is extracted from the investigation of a model with the quantum space
composed of an alternating sequence of these representations in \cite{jks}.

This study is devoted to an integrable vertex model built from
alternating sequences of the vector representation $W$ of $U_q\bigl(sl(2
\vert1)\bigr)$ and its dual $W^*$ in both horizontal and vertical direction.
In addition, the model allows for an inhomogeneity in the spectral
parameters. Within this arrangement, the north-west corner transfer matrix
(CTM)
is analyzed in the limit $q\to 0$. Even though this limit does not exist
for some elements of the R-matrix acting on $W\otimes W^*$ or $W^*
\otimes W$, the elements of the composite R-matrix defined on $W\otimes
W^*\otimes W\otimes W^*$ have a well defined limit.
Suitable boundary conditions provided, corner transfer matrices map
horizontal half-infinite sequences of vertical lattice links onto
vertical half-infinite sequences of horizontal lattice links or vice
versa. For
vertex model based on quantum affine algebras, the corner transfer matrix
is diagonal in the limit $q\to0$ \cite{kmn}.
Though the structure of the corner
transfer matrix elements remains nontrivial for the present model at $q=0$,
the simplification owed to this limit renders the spectrum
accessible.

For a variety of integrable models associated to quantum algebras, the
eigenvectors of the corner transfer matrices have been interpreted
as weight vectors of level-$k$ modules of the corresponding quantum
affine algebra \cite{ctm3}-\cite{jmo}. With these results, the concept
of vertex operators \cite{fr} leads to a mathematical description of
physical objects such as the transfer matrix or N-point correlators
\cite{dav}, \cite{jm}. These developments motivate the search for a
similar identification of the CTM-eigenvectors of the mixed
$U_q\bigl(sl(2\vert1)\bigr)$-model. To this aim,
the eigenvectors attributed to each CTM-eigenvalue greater than
$-5$ are compared to 
the weight states of the $U_q\bigl(\widehat{sl}(2\vert
1)\bigr)$-module $V(\Lambda_2)$ at level one with grade greater than $-5$. 
The eigenvalues of the generators of the Cartan subalgebra acting on these
eigenvectors via the infinitely folded coproduct are found
in one-to-one correspondence to the weights at a given grade 
identified with the CTM-eigenvalue. Furthermore, all eigenvectors
associated to nondegenerate CTM-eigenvalues are related to the weight states
of $V(\Lambda_2)$
with multiplicity one. Relying on these results,
a one-to-one correspondence between the CTM-eigenvectors
and the weight states of the $U_q\bigl(\widehat{sl}(2\vert1)
\bigr)_1$-module is conjectured. A more complete analysis as well as a
description  of the physical picture in terms of vertex operators will
be published subsequently.

The paper is organized as follows. Section \ref{sec:def} recalls
the definition of the quantum affine superalgebra $U_q\bigl(
\widehat{sl}(2\vert1)\bigr)$ and the various R-matrices related to the
vector representation of $U_q\bigl(sl(2\vert1)\bigr)$ and its dual.
Section \ref{sec:latt} specifies the integrable vertex model.
In section \ref{sec:plac}, the Boltzmann weights of the elementary
plaquettes composing the lattice model are evaluated in the limit $q\to0$.
The north-west corner transfer matrix for the inhomogeneous model is
introduced in first part of section \ref{sec:qzero}. Its spectrum is
investigated in subsections
\ref{sec:trigset}-\ref{sec:genset}. Section \ref{sec:mod}
regards the module $V(\Lambda)$ and its relation to the half-infinite
configurations on the lattice.
Some details relevant to subsections \ref{sec:nontrset} and
\ref{sec:genset} are provided in the appendix.

\section{The quantum affine superalgebra $U_q\bigl(\widehat{sl}(2\vert
1)\bigr)$}
\label{sec:def}

The quantum affine superalgebra
$U'_q\bigl(\widehat{sl}(2\vert1)\bigr)$ is the ${\bf C}$-algebra
generated by $\{e_i,f_i,q^{\pm h_i},\,i=0,1,2\}$ with the defining
relations
\bea
q^{h_i}\,q^{h_j}&=&q^{h_j}\,q^{h_i}\cr
\noalign{\medskip}
q^{h_i}\,e_j&=&q^{a_{ij}}\,e_j\,q^{h_i}\qquad\qquad q^{h_i}\,f_j=
q^{-a_{ij}}\,f_j\,q^{h_i}\cr
\noalign{\medskip}
[e_i,f_j]&=&\delta_{i,j}\,{q^{h_i}-q^{-h_i}\over q-q^{-1}}
\label{eq:def1}
\eea
and
\bea
f_0f_0f_i-[2]f_0f_if_0+f_if_0f_0&=&0\cr
\noalign{\medskip}
e_0e_0e_i-[2]e_0e_ie_0+e_ie_0e_0&=&0\qquad\mbox{for}\; i=1,2
\label{eq:def2}
\eea
\bea
[2]\bigl(f_1f_2f_1f_0f_2+f_2f_0f_1f_2f_1-f_2f_1f_2f_0f_1-f_1f_0f_2f_1f_2+
f_1f_2f_0f_2f_1-f_2f_1f_0f_1f_2\bigr)&+&\cr
\noalign{\bigskip}
+f_0f_1f_2f_1f_2-f_1f_2f_1f_2f_0-f_0f_2f_1f_2f_1+
f_2f_1f_2f_1f_0=0&\;&\cr
\noalign{\bigskip}
[2]\bigl(e_1e_2e_1e_0e_2+e_2e_0e_1e_2e_1-e_2e_1e_2e_0e_1-
e_1e_0e_2e_1e_2+
e_1e_2e_0e_2e_1-e_2e_1e_0e_1e_2\bigr)&+&\cr
\noalign{\bigskip}
+e_0e_1e_2e_1e_2-e_1e_2e_1e_2e_0-e_0e_2e_1e_2e_1+e_2e_1e_2e_1e_0=0&\;&
\label{eq:def3}
\eea
The super commutator in the last equation
of (\ref{eq:def1}) is defined by
\bea
[a,b]=ab-(-1)^{\vert a\vert\cdot
\vert b\vert}\,ba\qquad
\forall a,b\in U'_q\bigl(\widehat{sl}(2\vert1)\bigr)
\eea
with the ${\bf Z}_2$-grading $\vert\cdot\vert:U'_q\bigl(\widehat{sl}
(2\vert1)\bigr)\to {\bf Z}_2$ given by $\vert e_1\vert=\vert e_2\vert
=\vert f_1\vert=\vert f_2\vert=1$ and $\vert e_0\vert=\vert f_0\vert=
\vert q^{h_i}\vert=0$. Incorporating a generator $d$ with the property
\bea
[d,e_1]=\delta_{i,0}\,e_i\qquad[d,f_i]=-\delta_{i,0}
\,f_i\qquad[d,h_i]=0\qquad\mbox{for}\; i=0,1,2
\label{eq:ddef}
\eea
yields the quantum affine superalgebra $U_q\bigl(\widehat{sl}(2\vert1)
\bigr)$.
Choosing both classical simple roots of the superalgebra 
odd, the matrix elements of the symmetrized Cartan matrix
$a$ are $a_{00}=2,\;a_{11}=a_{22}=0,\;a_{12}=a_{21}=-a_{01}=
-a_{10}=-a_{02}=-a_{20}=1$. In terms of the basis $\bigl\{\tau_1,\tau_2,
\tau_3\bigr\}$ with the bilinear form $(\tau_i,\tau_j)=-(-1)^i\delta_{i,
j}$, the classical simple roots $\bar{\alpha}_i$ and the classical
weights $\bar{\Lambda}_i$ can be expressed
by $\bar{\alpha}_i=-(-1)^i(\tau_i+\tau_{i+1})$ and $\bar{\Lambda}_i=
\sum_{j=1}^i\tau_j-\delta_{i,1}\sum_{j=1}^3\tau_i$ with $i=1,2$.
An affine root $\delta$ and an affine
weight $\Lambda_0$ are introduced by $(\Lambda_0,\Lambda_0)=(\delta,
\delta)=(\Lambda_0,\tau_i)=(\delta,\tau_i)=0$ and $(\Lambda_0,\delta)=1$.
Then the set of simple roots $\{\alpha_i\}_{i=0,1,2}$ of
$U_q\bigl(\widehat{sl}(2\vert1)\bigr)$ is given by $\alpha_0=\delta-
\bar{\alpha}_1-\bar{\alpha}_2$ and $\alpha_i=\bar{\alpha}_i$ for
$i=1,2$. The remaining affine weights are $\Lambda_i=\bar{\Lambda}_i+
\Lambda_0$ with $i=1,2$. The free abelian group 
$P={\bf Z}\Lambda_0\oplus{\bf Z}\Lambda_1\oplus{\bf Z}\Lambda_2\oplus
{\bf Z}\delta$ is referred to as the weight lattice. Its dual lattice
$P^*={\bf Z}h_0\oplus{\bf Z}h_1\oplus{\bf Z}h_2\oplus{\bf Z}d$
may be identified with a subset of $P$ via $(,)$ setting $\alpha_i=h_i$
and $d=\Lambda_0$.

A graded Hopf algebra structure is provided by the coproduct
\bea
\Delta(e_i)=q^{h_i}\otimes e_i+e_i\otimes1\qquad\Delta(f_i)=f_i\otimes
q^{-h_i}+1\otimes f_i\qquad\Delta(q^{h_i})=q^{h_i}\otimes q^{h_i}
\label{eq:codef}
\eea
the antipode
\bea
S(e_i)=-q^{-h_i}\,e_i\qquad S(f_i)=-f_i\,q^{h_i}\qquad S(q^{h_i})=q^{-h_i}
\label{eq:antipdef}
\eea
and the counit
\bea
\epsilon(e_i)=\epsilon(f_i)=\epsilon(h_i)=0\qquad\epsilon(1)=1
\label{eq:coudef}
\eea
The Drinfeld basis of generators proves well suited for purposes related
to the bosonization of the superalgebra. In terms of this basis,
$U'_q\bigl(\widehat{sl}(2\vert1)\bigr)$ is generated by
$\{E^{i\pm}_m,H^i_n,q^{\pm h_i}\}$ with $i=1,2,\,m\in{\bf Z},n\in {\bf Z}
-\{0\}$ and the central elements $\gamma^{\pm{1\over2}}$ subject to
\bea
[H^i_n,H^j_m]&=&\delta_{n+m,0}\,{q^{a_{ij}n}-q^{-a_{ij}n}\over n(q-q^{-1})}
\,{\gamma^n-\gamma^{-n}\over q-q^{-1}}\cr
\noalign{\bigskip}
q^{h_j}\,E^{i\pm}_m\,q^{-h_j}&=&q^{\pm a_{ij}}\,E^{i\pm}_m\cr
\noalign{\bigskip}
[H^i_n,E^{j\pm}_m]&=&\pm\,{q^{a_{ij}n}-q^{-a_{ij}n}\over n(q-q^{-1})}\,
\gamma^{\mp{\vert n\vert\over2}}\,E^{j\pm}_{n+m}\cr
\noalign{\bigskip}
\Bigl[E^{i+}_n,E^{j-}_m\Bigr]
&=&\delta_{ij}\,{\gamma^{n-m\over2}\,\Psi^{i+}_{n+m}-
\gamma^{m-n\over2}\,\Psi^{i-}_{n+m}\over q-q^{-1}}
\label{eq:def4}
\eea
with
\bea
\sum_{n\ge0}\Psi^{i+}_nz^{-n}&=&q^{h_i}\,\exp\Bigl((q-q^{-1})\sum_{n>0}
H^i_nz^{-n}\Bigr)\cr
\noalign{\medskip}
\sum_{n\ge0}\Psi^{i-}_{-n}z^n&=&q^{-h_i}\,\exp\Bigl(-(q-q^{-1})\sum_{n>0}
H^i_{-n}z^n\Bigr)
\eea
and
\bea
&&\Bigl[E^{i\pm}_m,E^{i\pm}_n\Bigr]=0\qquad i=1,2\cr
\noalign{\bigskip}
&&E^{1\pm}_{n+1}E^{2\pm}_m+q^{\pm 1}\,E^{2\pm}_mE^{1\pm}_{n+1}=
q^{\pm1}\,E^{1\pm}_nE^{2\pm}_{m+1}+E^{2\pm}_{m+1}E^{1\pm}_n
\label{eq:def5}
\eea
The above choice of simple roots implies the
${\bf Z}_2$-grading $\vert E^{i\pm}_m\vert=1$ and
$\vert H^i_n\vert=\vert q^{\pm h_i}\vert=\vert\gamma\vert=0$.
The Drinfeld generators are related to the Chevalley basis (\ref{eq:def1})
-(\ref{eq:def3}) by
\bea
e_i&=&E^{i+}_0\qquad f_i=E^{i-}_0\qquad\mbox{for}\;i=1,2\cr
\noalign{\medskip}
e_0&=&\Bigl(E^{2-}_0E^{1-}_1+q\,E^{1-}_1E^{2-}_0\Bigr)\,q^{-h_1-h_2}\cr
\noalign{\medskip}
f_0&=&-q^{h_1+h_2}\,\Bigl(E^{1+}_{-1}E^{2+}_0+q^{-1}\,E^{2+}_0E^{1+}_{-1}
\Bigr)
\label{eq:rel1}
\eea
A three-dimensional module $W$ of $U'_q\bigl(\widehat{sl}(2\vert1)\bigr)$
with basis $\bigl\{w_i\bigr\}_{0\leq i\leq2}$ is given by 
\bea
\begin{array}{llll}
f_0\,w_2&=qw_0\qquad&e_0\,w_0&=q^{-1}w_2\cr
\noalign{\bigskip}
f_1\,w_0&=w_1\qquad&e_1\,w_1&=w_0\cr
\noalign{\bigskip}
f_2\,w_1&=w_2\qquad&e_2\,w_2&=-w_1
\end{array}
\eea
and
\bea
\begin{array}{llllll}
h_0\,w_0&=-w_0\qquad&h_1\,w_0&=w_0\qquad&h_2\,w_0&=0\cr
\noalign{\bigskip}
h_0\,w_1&=0\qquad&h_1\,w_1&=w_1\qquad&h_2\,w_1&=-w_1\cr
\noalign{\bigskip}
h_0\,w_2&=w_2\qquad&h_1\,w_2&=0\qquad&h_2\,w_2&=-w_2
\end{array}
\eea
The ${\bf Z}_2$-grading on $W$ is fixed by
$\vert w_0\vert=\vert w_2\vert=0$ and $\vert w_1\vert=1$.
Given an anti-automorphism $\phi$ of the superalgebra, the dual space
of $W$ endowed with $U'_q\bigl(\widehat{sl}(2\vert1)\bigr)$-structure
\bea
\langle xw^*\vert w\rangle=(-1)^{\vert x\vert\cdot\vert w^*\vert}\;
\langle w^*\vert \phi(x)w\rangle\qquad\forall x\in U'_q\bigl(\widehat{sl}(2
\vert1)\bigr)
\eea
is usually denoted by $W^{*\phi}$. In the following, $W^{*S}$ will be
denoted by $W^*$ for brevity. The basis $\bigl\{w^*_i\bigr\}_{0\leq i
\leq 2}$ of $W^*$ with $\vert w^*_0\vert=\vert w^*_2\vert=0$ and $\vert
w^*_1\vert=1$
may be chosen such that the action of the superalgebra
reads
\bea
\begin{array}{llll}
f_1\,w^*_1&=qw^*_0\qquad&e_1\,w^*_0&=-q^{-1}w^*_1\cr
\noalign{\bigskip}
f_2\,w^*_2&=-q^{-1}w^*_1\qquad&e_2\,w_1^*&=-qw^*_2\cr
\noalign{\bigskip}
f_0\,w_0^*&=-q^2w_2^*\qquad&e_0\,w_2^*&=-q^{-2}w_0^*
\end{array}
\eea
\bea
\begin{array}{llllll}
h_0\,w^*_0&=w^*_0\qquad&h_1\,w^*_0&=-w^*_0\qquad&h_2\,w^*_0&=0\cr
\noalign{\bigskip}
h_0\,w^*_1&=0\qquad&h_1\,w^*_1&=-w^*_1\qquad&h_2\,w^*_1&=w^*_1\cr
\noalign{\bigskip}
h_0\,w^*_2&=-w^*_2\qquad&h_1\,w^*_2&=0\qquad&h_2\,w^*_2&=w^*_2
\end{array}
\eea
A $U_q\bigl(\widehat{sl}(2\vert1)\bigr)$-structure can be implemented on
the evaluation modules $W_z=W\otimes F[z,z^{-1}]$ and $W^*_z=W^*\otimes
F[z,z^{-1}]$ via
\bea
\begin{array}{llll}
e_i\bigl(v_j\otimes z^n\bigr)&=e_iv_j\otimes z^{n+\delta_{i,0}}\qquad&
f_i\bigl(v_j\otimes z^n\bigr)&=f_iv_j\otimes z^{n-\delta_{i,0}}\cr
\noalign{\bigskip}
h_i\bigl(v_j\otimes z^n\bigr)&=h_iv_j\otimes z^n\qquad&d\bigl(v_j\otimes
z^n\bigr)&=n\,v_j\otimes z^n
\end{array}
\label{eq:evmodef}
\eea
with $i,j=0,1,2$ and  $v_j=w_j$ or $v_j=w^*_j$.

For two evaluation modules $V^{(m)}_{z_m}=W_{z_m}$ or $V^{(m)}_{z_m}=
W^*_{z_m}$, the
R-matrix $R(z_1/z_2)\in End\bigl(V^{(1)}_{z_1}\otimes V^{(2)}_{z_2}
\bigr)$ intertwines
the action of $U_q\bigl(\widehat{sl}(2\vert1)\bigr)$ according to
\bea
R(z_1/z_2)\,\Delta(x)=\Delta'(x)\,R(z_1/z_2)\qquad\forall x\in U_q\bigl(
\widehat{sl}(2\vert1)\bigr)
\label{eq:int}
\eea
where $\Delta'=\sigma\circ\Delta$ and $\sigma(x\otimes y)=(-1)^{\vert x
\vert\cdot\vert y\vert}\,y\otimes x$. In the remainder, subscripts will
indicate the choices of the evaluation modules where required for clarity.
The corresponding R-matrix elements are introduced by
\bea
R_{WW}(z_1/z_2)\bigl(w_i\otimes w_j\bigr)&=&\sum_{k,l=0,1,2}R_{i,j}^{k,l}
(z_1/z_2)\,w_k\otimes w_l\cr
\noalign{\bigskip}
R_{W^*W^*}(z_1/z_2)\bigl(w^*_i\otimes w^*_j\bigr)&=&\sum_{k,l=0,1,2}
R_{i^*,j^*}^{k^*,l^*}(z_1/z_2)\,w^*_k\otimes w^*_l\cr
\noalign{\bigskip}
R_{WW^*}(z_1/z_2)\bigl(w_i\otimes w^*_j\bigr)&=&\sum_{k,l=0,1,2}
R_{i,j^*}^{k,l^*}(z_1/z_2)\,w_k\otimes w^*_l\cr
\noalign{\bigskip}
R_{W^*W}(z_1/z_2)\bigl(w^*_i\otimes w_j\bigr)&=&\sum_{k,l=0,1,2}
R_{i^*,j}^{k^*,l}(z_1/z_2)\,w^*_k\otimes w_l
\label{eq:rmel}
\eea
Up to a scalar multiple, the R-matrix elements (\ref{eq:rmel})
are uniquely determined by the intertwining property
(\ref{eq:int}). For $V^{(1)}_{z_1}=W_{z_1}$ and
$V^{(2)}_{z_2}=W_{z_2}$, a solution of (\ref{eq:int}) is given by 
\bea
R_{0,0}^{0,0}(z)&=&R^{2,2}_{2,2}(z)=1\qquad\qquad R^{1,1}_{1,1}(z)=
{q^2-z\over1-q^2z}\cr
\noalign{\bigskip}
R^{i,j}_{i,j}(z)&=&{q(1-z)\over1-q^2z}\qquad i\neq j\cr
\noalign{\bigskip}
R^{j,i}_{i,j}(z)&=&-{(q^2-1)z\over1-q^2z}\qquad i<j\cr
\noalign{\bigskip}
R^{j,i}_{i,j}(z)&=&-{q^2-1\over1-q^2z}\qquad i>j
\label{eq:rsol}
\eea
The solution satisfies the initial condition
\bea
R_{i,j}^{k,l}(1)=R_{i^*,j^*}^{k^*,l^*}(1)=\delta_{i,l}
\delta_{j,k}\,(-1)^{\vert k\vert\cdot\vert l\vert}
\label{eq:init1}
\eea
Relations between R-matrix elements with respect to the various choices
of evaluation modules are conveniently stated introducing
\bea
\begin{array}{llll}
\bar R_{i,j}^{k,l}(z)&=(-1)^{\vert k\vert\cdot\vert l\vert}\,R_{i,j}^
{k,l}(z)\qquad&
\bar R_{i^*,j^*}^{k^*,l^*}(z)&=(-1)^{\vert k\vert\cdot\vert l\vert}\,
R_{i^*,j^*}^{k^*,l^*}(z)\cr
\noalign{\bigskip}
\bar R_{i^*,j}^{k^*,l}(z)&=(-1)^{\vert k\vert\cdot\vert l\vert}\,
R_{i^*,j}^{k^*,l}(z)\qquad&
\bar R_{i,j^*}^{k,l^*}(z)&=(-1)^{\vert k\vert\cdot\vert l\vert}\,
R_{i,j^*}^{k,l^*}(z)
\end{array}
\label{eq:relmod}
\eea
Then solutions to the intertwining condition (\ref{eq:int}) are
obtained from (\ref{eq:rsol}), (\ref{eq:relmod}) and 
\bea
\bar R^{k^*,l^*}_{i^*,j^*}(z)&=&\bar R^{i,j}_{k,l}(z)\cr
\noalign{\bigskip}
\bar R^{k,l^*}_{i,j^*}(z)&=&(-1)^{\vert k\vert-\vert i\vert}\;
\bar R^{j,k}_{l,i}\bigl(1/q^2z\bigr)\cr
\noalign{\bigskip}
\bar R^{k^*,l}_{i^*,j}(z)&=&
\bar R^{l,i}_{j,k}\bigl(1/z\bigr)
\label{eq:rrel}
\eea
In terms of the R-matrix elements (\ref{eq:rmel}), the second inversion
relation reads
\bea
R^{k,l}_{i,j}(z)=R^{j,i}_{l,k}(z)\qquad\qquad R^{k^*,l^*}_{i^*,j^*}(z)=
R^{j^*,i^*}_{l^*,k^*}(z)\qquad
\qquad
R^{k^*,l}_{i^*,j}(z)=R^{j,i^*}_{l,k^*}\bigl(z/q^2\bigr)
\label{eq:sym1}
\eea
A further symmetry of the R-matrix is stated by
\bea
\begin{array}{llll}
\bar R^{k,l}_{i,j}(z)&=z^{{1\over2}(k+\delta_{k,1}-i-\delta_{i,1})}\;
\bar R^{l,k}_{j,i}(z)\qquad&\bar R^{k^*,l^*}_{i^*,j^*}(z)&=
z^{-{1\over2}(k+\delta_{k,1}-i-\delta_{i,1})}\;\bar R^{l^*,k^*}_{j^*,
i^*}(z)\cr
\noalign{\bigskip}
\bar R^{k,l^*}_{i,j^*}(z)&=\bigl(q^2z\bigr)^{{1\over2}(k+\delta_{k,1}-i
-\delta_{i,1})}\;\bar R^{l,k^*}_{j,i^*}(z)
\qquad&\bar R^{k^*,l}_{i^*,j}(z)&=z^{-{1\over2}(k+\delta_{k,1}-i-\delta_
{i,1})}\;\bar R^{l^*,k}_{j^*,i}(z)
\end{array}
\label{eq:sym2}
\eea

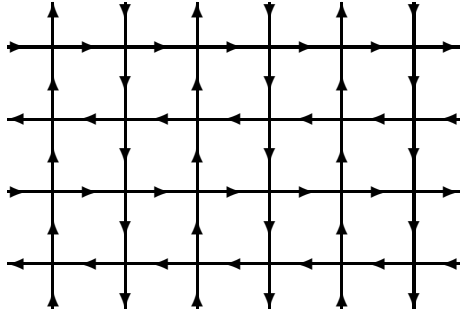
\section{The lattice model}
\label{sec:latt}

A section of the infinite lattice model considered in the following sections
is illustrated in Fig. 1. The modules $W$ and $W^*$ are associated in an
alternating sequence to the horizontal and vertical lines as indicated
by arrows pointing right or upwards for $W$ and pointing left or
downwards for $W^*$. Hence, the lattice model may be decomposed into
elementary plaquettes shown in Fig. 2. For each vertex, the matrix
elements (\ref{eq:relmod}) provide Boltzmann weights depending on the
configuration of basis elements $\bigl\{w_i\bigr\}_{0\leq i\leq2}$
or $\bigl\{w^*_i\bigr\}_{0\leq i\leq2}$ attributed to the adjacent
links and on the spectral parameters chosen for the type of
vertex as specified in the left part of Fig. 2.
\vskip 1cm
\begin{center}
\setlength{\unitlength}{0.8cm}
\begin{picture}(7.4,5)
\thicklines
\put(0,1.95){\vector(1,0){0.3}}
\put(0,4.35){\vector(1,0){0.3}}
\multiput(0.3,1.95)(1.2,0){6}{\vector(1,0){1.2}}
\multiput(0.3,4.35)(1.2,0){6}{\vector(1,0){1.2}}
\put(7.5,0.75){\vector(-1,0){0.3}}
\put(7.5,3.15){\vector(-1,0){0.3}}
\multiput(7.2,0.75)(-1.2,0){6}{\vector(-1,0){1.2}}
\multiput(7.2,3.15)(-1.2,0){6}{\vector(-1,0){1.2}}
\multiput(0.75,0)(2.4,0){3}{\vector(0,1){0.3}}
\multiput(0.75,0.3)(2.4,0){3}{\vector(0,1){1.2}}
\multiput(0.75,1.5)(2.4,0){3}{\vector(0,1){1.2}}
\multiput(0.75,2.7)(2.4,0){3}{\vector(0,1){1.2}}
\multiput(0.75,3.9)(2.4,0){3}{\vector(0,1){1.2}}
\multiput(1.95,5.1)(2.4,0){3}{\vector(0,-1){0.3}}
\multiput(1.95,4.8)(2.4,0){3}{\vector(0,-1){1.2}}
\multiput(1.95,3.6)(2.4,0){3}{\vector(0,-1){1.2}}
\multiput(1.95,2.4)(2.4,0){3}{\vector(0,-1){1.2}}
\multiput(1.95,1.2)(2.4,0){3}{\vector(0,-1){1.2}}
\end{picture}\par
\vskip 0.4cm
Fig. 1: The lattice model
\end{center}

\begin{center}
\setlength{\unitlength}{1cm}
\begin{picture}(10,6)
\thicklines
\put(0.8,1.4){\line(1,0){0.1}}
\put(2.1,1.4){\vector(-1,0){1.2}}
\put(3.3,1.4){\vector(-1,0){1.2}}
\put(3.7,1.4){\vector(-1,0){0.4}}
\put(0.8,3.2){\vector(1,0){0.4}}
\put(1.2,3.2){\vector(1,0){1.2}}
\put(2.4,3.2){\vector(1,0){1.2}}
\put(3.6,3.2){\line(1,0){0.1}}
\put(1.35,0.9){\vector(0,1){0.4}}
\put(1.35,1.3){\vector(0,1){1.2}}
\put(1.35,2.5){\vector(0,1){1.2}}
\put(1.35,3.7){\line(0,1){0.1}}
\put(3.15,3.8){\vector(0,-1){0.45}}
\put(3.15,3.4){\vector(0,-1){1.2}}
\put(3.15,2.2){\vector(0,-1){1.25}}
\put(3.15,1){\line(0,-1){0.1}}
\put(1.5,1.6){$wz$}
\put(2.2,2.8){$w^{-1}z$}
\put(1.5,2.8){$z$}
\put(2.85,1.6){$z$}
\put(0.7,0){Spectral parameters}

\put(6.8,1.4){\line(1,0){0.1}}
\put(8.1,1.4){\vector(-1,0){1.2}}
\put(9.3,1.4){\vector(-1,0){1.2}}
\put(9.7,1.4){\vector(-1,0){0.4}}
\put(6.8,3.2){\vector(1,0){0.4}}
\put(7.2,3.2){\vector(1,0){1.2}}
\put(8.4,3.2){\vector(1,0){1.2}}
\put(9.6,3.2){\line(1,0){0.1}}
\put(7.35,0.9){\vector(0,1){0.4}}
\put(7.35,1.3){\vector(0,1){1.2}}
\put(7.35,2.5){\vector(0,1){1.2}}
\put(7.35,3.7){\line(0,1){0.1}}
\put(9.15,3.8){\vector(0,-1){0.45}}
\put(9.15,3.4){\vector(0,-1){1.2}}
\put(9.15,2.2){\vector(0,-1){1.25}}
\put(9.15,1){\line(0,-1){0.1}}
\put(6.45,3.15){$i$}
\put(6.4,1.35){$j^*$}
\put(7.25,0.5){$k$}
\put(9.05,0.5){$l^*$}
\put(7.27,4){$i'$}
\put(9,4){$j'^*$}
\put(9.95,3.15){$k'$}
\put(9.9,1.35){$l'^*$}
\put(6.5,0){Assignment of indices}
\end{picture}\par
\vskip 0.7cm
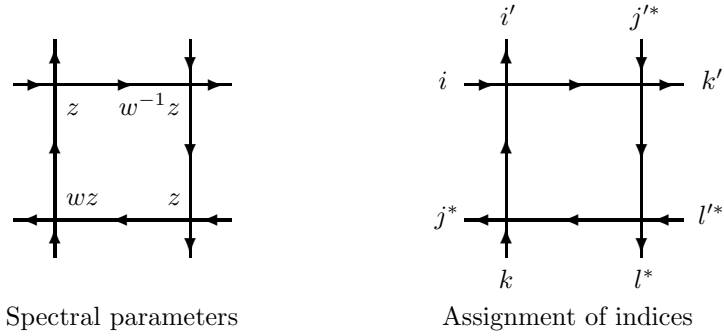
Fig. 2: The elementary plaquette
\end{center}
The present investigation applies to any finite value $w$ specifying the
inhomogeneity in the spectral parameters. An analysis of the limits
$w\to0$ and $w\to\infty$, where the R-matrices on $W\otimes W^*$ and
$W^*\otimes W$ tend towards their braid limits, will be presented
separately.

With the assignment of indices shown in the right part of Fig. 2,
the Boltzmann weight corresponding to an elementary plaquette is
given by
\bea
R_{\,i,\;j^*;\;k,\;l^*}^{i',j'^*;k',l'^*}(w,z)=\sum_{\tilde j,\tilde k,
\tilde l,\tilde n=0,1,2}\bar R_{\,\tilde n,\;\tilde l^*}^{\,k',j'^*}
(w^{-1}z)\;\bar R_{\,\tilde j^*,l^*}^{\,l'^*,\tilde l^*}(z)\;
\bar R_{\,i,\tilde k}^{\,\tilde n,i'}(z)\;
\bar R_{\,j^*,k}^{\,\tilde j^*,\tilde k}(wz)
\label{eq:plbo}
\eea
Use of the solutions (\ref{eq:rsol})-(\ref{eq:rrel}) of the intertwining
condition (\ref{eq:int}) for the matrix elements in the last equation
yields an integrable vertex model. Due to the initial condition
(\ref{eq:init1})
and the unitarity relation $\sum_{k,l=0,1,2}\bar R^{j',i'^*}_{
\,l,\;k^*}(w^{-1})\bar R_{i^*,j}^{k^*,l}(w)=\delta_{i,i'}\delta_{j,j'}$ the
matrix elements of $R(w,z):\bigl(\otimes W\otimes W^*\bigr)^2
\longrightarrow\bigl(\otimes W\otimes W^*\bigr)^2$ satisfy
\bea
R_{\,i,\;j^*;\;k,\;l^*}^{i',j'^*;k',l'^*}(w,1)=\delta_{i,i'}\delta_{j,j'}
\delta_{k,k'}\delta_{l,l'}
\label{eq:init2}
\eea
The property (\ref{eq:init2}) may be viewed as initial condition for
$R(w,z)$.

\section{The limit $q\to0$}
\label{sec:plac}

The Boltzmann weights of homogeneous vertex models based on
finite-dimensional representations of quantum algebras simplify
drastically in the limit $q\to0$ \cite{kmn}. A remarkable simplification
occurs in the present case, too. Well-defined limits are found for
the Boltzmann weights (\ref{eq:plbo}) of the elementary plaquettes
even though this is not the case for all single R-matrix elements
(\ref{eq:rrel}). Keeping $z$ and $w$ fixed, the matrix elements
$R^{i',j'^*;k',l'^*}_{\,i,\;j^*;\;k,\;l^*}(qw,z)$ tend to values
independent of $w$ when the limit $q\to0$ is performed:
\bea
\lim_{q\to0}R^{i',j'^*;k',l'^*}_{\,i,\;j^*;\;k,\;l^*}(qw,z)\equiv
P^{i',j'^*;k',l'^*}_{\;i,\;j^*\,;\;k,\;l^*}(z)
\label{eq:limdef}
\eea
Use of the explicit expressions (\ref{eq:rsol}) in
(\ref{eq:rrel}) yields the following results for these limits :
\bea
P^{i,j^*;k,l^*}_{\,i,j^*;k,l^*}(z)=z^{-y_{i,j,k}-y_{l,k,j}}
\label{eq:lim1a}
\eea
with
\bea
\begin{array}{lll}
y_{i,j,k}&=0\qquad&i<k\cr
\noalign{\bigskip}
y_{i,j,k}&=1\qquad&i>k,\;j\neq0\;\mbox{for}\;i=1\cr
\noalign{\bigskip}
y_{1,0,0}&=0\qquad&\cr
\noalign{\bigskip}
y_{i,j,k}&=0\qquad&i=k=1\cr
\noalign{\bigskip}
y_{i,j,k}&=1\qquad&i=k=0,2,\;\;j\neq i\cr
\noalign{\bigskip}
y_{i,j,k}&=0\qquad&i=j=k=0,2
\end{array}
\label{eq:lim1b}
\eea
and
\bea
P^{i,(1^*;1+0^*;0),l^*}_{\;i,(1^*;1-0^*;0),l^*}(z)&=&0\cr
\noalign{\bigskip}
P^{i,(1^*;1-0^*;0),l^*}_{\;i,(1^*;1+0^*;0),l^*}(z)&=&2(\delta_{i,1}
+\delta_{l,1})\cdot{z-1\over z^{1+\delta_{i,2}+\delta_{l,2}}}
\qquad\forall i,l
\label{eq:lim2}
\eea
In addition, the following nondiagonal matrix elements of $P(z)$
are found:
\bea
P^{k,1^*;1,l^*}_{\;1,1^*;k,l^*}(z)&=&-P^{k,0^*;0,l^*}_{\;
1,1^*;
k,l^*}(z)={z-1\over z^{1+\delta_{l,2}}}\qquad\mbox{for}\;k\neq1\cr
\noalign{\bigskip} 
P^{i,1^*;1,j^*}_{\;i,j^*;1,1^*}(z)&=&-P^{i,0^*;0,j^*}_{\;i,
j^*;1,1^*}(z)={z-1\over z^{1+\delta_{i,2}}}\qquad\mbox{for}\;j\neq1
\label{eq:lim3}
\eea
\bea
P^{k,1^*;1,l^*}_{\;2,2^*;k,l^*}
(z)&=&-P^{k,0^*;0,l^*}_{\;2,2^*;k,l^*}(z)=-{z^{1+
\delta_{k,2}\delta_{l,2}}-1\over z^{1+\delta_{l,2
}}}\qquad\qquad P^{i,1^*;1,j^*}_{\;i,j^*;2,2^*}(z)=-P^{i,0^*;0,
j^*}_{\;i,j^*;2,2^*}(z)=-{z^{1+\delta_{i,2}\delta_{j,2}}-1\over
z^{1+\delta_{i,2}}}
\label{eq:lim4}
\eea
and
\bea
P^{1,1^*;i,l^*}_{\;i,1^*;1,l^*}(z)+P^{1,1^*;i,l^*}_{\;i,0^*;0,l^*}
(z)&=&2\,{z-1
\over z^{1+\delta_{l,2}}}\qquad i\neq1\cr
\noalign{\bigskip}
P^{i,l^*;1,1^*}_{\;i,1^*;1,l^*}(z)+P^{i,l^*;1,1^*}_{\;i,0^*;0,l^*}(z)
&=&2\,{z-1\over z^{1+\delta_{i,2}}}\qquad l\neq1\cr
\noalign{\bigskip}
P^{2,2^*;i,l^*}_{\;i,1^*;1,l^*}(z)+P^{2,2^*;i,l^*}_{\;i,0^*;0,l^*}(z)
&=&2\,{z^{1+\delta_{i,2}\delta_{l,2}}-1\over z^{1+\delta_{l,2}}}\cr
\noalign{\bigskip}
P^{i,l^*;2,2^*}_{\;i,1^*;1,l^*}(z)+P^{i,l^*;2,2^*}_{\;i,0^*;0,l^*}(z)
&=&2\,{z^{1+\delta_{i,2}\delta_{l,2}}-1\over z^{1+\delta_{i,2}}}
\label{eq:lim5}
\eea
All other matrix elements of $P(z)$ vanish. In contrast to the cases
studied in \cite{kmn}, the matrix of Boltzmann weights does not
assume a diagonal form in the limit of vanishing $q$. However,
the simplification proves sufficient to establish a link to the
representation theory of the affine superalgebra. As in the case
of lattice models related to finite-dimensional representations
of quantum affine algebras \cite{ctm3}-\cite{jmo},
this link is provided by the corner transfer matrices of the model.

\section{The corner transfer matrix in the limit $q\to 0$}
\label{sec:qzero}

\subsection{The inhomogeneous corner transfer matrix}
\label{sec:ctm}

Corner transfer matrices may be introduced for the present model in
close analogy to the construction developed for the eight-vertex model
in \cite{bax1,bax2}. A triangular subsection $A_N$ of the upper left
quadrant is considered. Its vertical(horizontal) boundaries coincide
with $2N+1$ horizontal(vertical) links on the boundaries of the quadrant.
For a fixed configuration of basis elements $\bigl\{w_i\bigr\}_{0\leq i
\leq2}$ or $\bigl\{w^*_i\bigr\}_{0\leq i\leq2}$ on all links of its
diagonal boundary, the Boltzmann weights of the subsection $A_N$ yield
a map
\bea
A^{(N)}(w,z):\;W^*\bigl(\otimes W\otimes W^*\bigr)^{N+1}\longrightarrow
W^*\bigl(\otimes W\otimes W^*\bigr)^{N+1}
\label{eq:map1}
\eea
At $z=1$, (\ref{eq:map1}) reduces to the identity map due to the initial
conditions (\ref{eq:init1}) and (\ref{eq:init2}). Thus a Hamiltonian
corresponding to the section $A_N$ can be introduced by
\bea
h^{(N)}_{CTM}(w)\equiv(N+1)h_{2N+2,2N+1}+\sum_{\tilde N=1}^N\tilde N\,
h_{2(\tilde N+1),2\tilde N+1;\;2\tilde N,2\tilde N-1}(w)
\label{eq:hctm1}
\eea
In (\ref{eq:hctm1}), $h_{2(\tilde N+1),2\tilde N+1;\;2\tilde N,2\tilde N-1}
(w)$ denotes the operator
$h(w)$ acting on the $(2\tilde N-1)$-th to the $(2(\tilde N+1))$-th
component of
$\bigl(\otimes W\otimes W^*\bigr)^{N+1}$ counted from the right, with
$h(w)$ defined by the expansion
\bea
R(w,z)={\bf 1}+(z-1)\,h(w)+O\bigl((z-1)^2\bigr)
\label{eq:hdef1}
\eea
Similarly, $h_{2N+2,2N+1}$ is the operator $h$ defined by
$\sigma\,R_{W^*W^*}(z)={\bf 1}+(z-1)\,h+O\bigl((z-1)^2\bigr)$
acting on the two leftmost components of $W^*\bigl(\otimes W\otimes W^*
\bigr)^{N+1}$. The boundary condition imposed on the diagonal boundary
of $A_N$ has to be chosen consistent with the expansion of $R(w,z)$
and $\sigma R_{W^*W^*}(z)$ at $z=1$. With respect to a particular boundary
condition, the large-$N$ limit of
$h^{(N)}_{CTM}(w)$ is referred to as the corner transfer matrix
Hamiltonian $h_{CTM}(w)$. The boundary condition adopted in the following
is specified by attributing only $w_2$ or $w^*_2$ to any
link on the diagonal boundary. In the remainder of this section,
$h^{(N)}_{CTM}(w)$ will be investigated in the limit $q\to0$ and
$N\to \infty$. The corresponding operator is denoted by
\bea
H_0=\lim_{N\to\infty}\lim_{q\to0}h^{(N)}_{CTM}(w)
\label{eq:hlimdef}
\eea
Examination of (\ref{eq:lim1a})-(\ref{eq:lim5}) shows that $H_0$
does not depend on the (finite) value  of $w$.

\subsection{A restricted set of configurations}
\label{sec:trigset}

A particular configuration $\bigl(\ldots\otimes w_{i_2}\otimes w^*_{j_2}
\otimes w_{i_1}\otimes w^*_{j_1}\bigr)$ on the horizontal or vertical
boundary of the corner transfer matrix may be abbreviated writing
\bea
\bigl(\ldots,j_2^*i_2,j_1^*i_1,j_0^*\bigr)\equiv\bigl(\ldots\otimes
w^*_{j_2}\otimes w_{i_2}\otimes w^*_{j_1}\otimes w_{i_1}
\otimes w^*_{j_0}\bigr)
\label{eq:state}
\eea
According to the boundary condition fixed above, only finitely many
$i_n$, $j_n$ differ from $2$.
For a large subset of configurations (\ref{eq:state})
to be specified in this subsection, the
matrix elements of the corner transfer Hamiltonian can be arranged
in a trigonal form.

Making use of (\ref{eq:lim1a}) and (\ref{eq:lim4}) in
(\ref{eq:hctm1})-(\ref{eq:hlimdef}) and of (\ref{eq:rsol})-(\ref{eq:rrel}),
the action of $H_0$ on the configuration
$\bigl(\ldots,2^*2,2^*2,i^*\bigr)$ with $i=0,1,2$ is easily evaluated:
\bea
H_0\bigl(\ldots,2^*2,2^*2,i^*\bigr)=&-&
(2-\delta_{i,0}\delta_{i,1})\bigl(\ldots,2^*2,2^*2,(1^*1-0^*0),i^*\bigr)
\cr
\noalign{\bigskip}
&-&2\Bigl\{2\bigl(\ldots,
2^*2,2^*2,(1^*1-0^*0),2^*2,i^*\bigr)
+3\bigl(\ldots,2^*2,(2^*2,1^*1-0^*0),2^*2,2^*2,i^*
\bigr)+\ldots\Bigr\}
\label{eq:h1}
\eea
The configurations on the rhs of (\ref{eq:h1}) are obtained from $\bigl(
\ldots,2^*2,2^*2,i^*\bigr)$ by replacing one subsequence $2^*2$ by
the difference $1^*1-0^*0$.
Taking into account (\ref{eq:lim1a}), (\ref{eq:lim1b}) and (\ref{eq:lim2}),
the action of $H_0$ on these configurations is found:
\bea
&&H_0\bigl(\ldots,2^*2,2^*2,(1^*1-0^*0),(2^*2)^n,i^*\bigr)=-\bigl(2(n+1)-
\delta_{n,0}(\delta_{i,0}+\delta_{i,1})\bigr)\,
\bigl(\ldots,2^*2,2^*2,(1^*1-0^*0),(2^*2)^n,i^*\bigr)\cr
\noalign{\bigskip}
&&\qquad\qquad
-\sum_{m=0}^{n-2}(m+1)
(2-\delta_{m,0}\,\delta_{i,0}\delta_{i,1})\,\bigl(\ldots,2^*2,2^*2,
\bigl(1^*1-0^*0),(2^*2)^{n-m-1},(1^*1-0^*0),(2^*2)^{m},i^*\bigr)\cr
\noalign{\bigskip}
&&\qquad\qquad-(n-\delta_{n,1}\,\delta_{i,0}\delta_{i,1})
\bigl(\ldots,2^*2,2^*2,(1^*1-0^*0)^2,(2^*2)^{n-1},i^*
\bigr)\cr
\noalign{\bigskip}
&&\qquad\qquad
-(n+2)\bigl(\ldots,2^*2,2^*2,(1^*1-0^*0)^2,(2^*2)^n,i^*\bigr)\cr
\noalign{\bigskip}
&&\qquad\qquad
-2\Bigl\{(n+3)\bigl(\ldots,2^*2,2^*2,(1^*1-0^*0),2^*2,(1^*1-0^*0),
(2^*2)^n,i^*\bigr)+\cr
\noalign{\bigskip}
&&\qquad\qquad\qquad
+(n+4)\bigl(\ldots,2^*2,2^*2,(1^*1-0^*0,)2^*2,2^*2,(1^*1-0^*0),
(2^*2)^n,i^*\bigr)+\cr
\noalign{\bigskip}
&&\qquad\qquad\qquad
+(n+5)\bigl(\ldots,2^*2,2^*2,(1^*1-0^*0),2^*2,2^*2,2^*2,(1^*1-
0^*0),(2^*2)^n,i^*\bigr)+\ldots\Bigr\}
\label{eq:h2}
\eea
With the second line dropped for $n=0,1$, equation (\ref{eq:h2}) is
valid for all $n\geq0$. Inspection of (\ref{eq:lim1a})-(\ref{eq:lim5})
allows for a description of the repeated action of $H_0$ on the
configurations on the rhs of (\ref{eq:h1}), (\ref{eq:h2}).
To facilitate notation of the explicit expressions it is useful to
introduce the abbreviation
\bea
\tau\bigl\{k_{t_l}^{(l)}\bigr\}_{1\leq t_l\leq s_l}\equiv\,,(1^*1-0^*0)^{
k^{(l)}_{1}},2^*2,(1^*1-0^*0)^{k^{(l)}_{2}},2^*2,(1^*1-0^*0)^{k^{
(l)}_{3}},
\ldots,2^*2,(1^*1-0^*0)^{k^{(l)}_{s_l}},
\eea
with $s_l=1,2,3,\ldots$ and $k^{(l)}_{t_l}=1,2,3,\ldots$ for $1\leq t\leq
s_l$. Any configuration emerging from repeated action of $h_0$ on
$\bigl(\ldots,2^*2,2^*2,i^*\bigr)$ can be written
\bea
\Bigl(\ldots,2^*2,2^*2\,\tau\bigl\{k^{(R)}_{t_R}\bigr\}_{1\leq t_R\leq
s_R}(2^*2)^{n_R}\tau\bigl\{k^{(R-1)}_{t_{R-1}}\bigr\}_{1\leq t_{R-1}\leq
s_{R-1}}(2^*2)^{n_{R-1}},\ldots,(2^*2)^{n_2}\tau\bigl\{k^{(1)}_{t_1}
\bigr\}_{1\leq t_1\leq s_1}(2^*2)^{n_1},i^*\Bigr)
\label{eq:conf1}
\eea
with $n_1\geq0$ and $n_l\geq2$ for $l>1$. A configuration of the form
(\ref{eq:conf1}) is composed
from $r=\sum_{l=1}^R\sum_{t_l=1}^{s_l}k^{(l)}_{t_l}$ subsequences
$1^*1-0^*0$ placed between subsequences $2^*2$ and the right end $\ldots,
i^*\bigr)$. The action of $H_0$ on (\ref{eq:conf1}) is given by
\bea
&&H_0\biggl(\Bigl(\ldots,2^*2,\,\tau\bigl\{k^{(R)}_{t_R}\bigr\}_{1\leq
t_R\leq
s_R}(2^*2)^{n_R}\tau\bigl\{k^{(R-1)}_{t_{R-1}}\bigr\}_{1\leq t_{R-1}\leq
s_{R-1}}(2^*2)^{n_{R-1}},\ldots,(2^*2)^{n_2}\tau\bigl\{k^{(1)}_{t_1}
\bigr\}_{1\leq t_1\leq s_1}(2^*2)^{n_1},i^*\Bigr)\biggr)\cr
\noalign{\bigskip}
&&=-\alpha\bigl(\bigl\{k^{(l)}_{t_l}\bigr\}_{1\leq t_l\leq s_l,\,1\leq
l\leq R},\;\{n_l\}_{1\leq l\leq R}\bigr)\cdot\cr
\noalign{\bigskip}
&&\qquad\Bigl(\ldots,2^*2,\,\tau\bigl\{k^{(R)}_{t_R}\bigr\}_{1\leq t_R\leq
s_R}(2^*2)^{n_R}\tau\bigl\{k^{(R-1)}_{t_{R-1}}\bigr\}_{1\leq t_{R-1}\leq
s_{R-1}}(2^*2)^{n_{R-1}},\ldots,(2^*2)^{n_2}\tau\bigl\{k^{(1)}_{t_1}
\bigr\}_{1\leq t_1\leq s_1}(2^*2)^{n_1},i^*\Bigr)\cr
\noalign{\bigskip}
&&-\sum_{m_1=0}^{n_1-1}(2-\delta_{m_1,n_1-1}-\delta_{m_1,0}\,
\delta_{i,0}\delta_{i,1})\,(m_1+1)\cdot\cr
\noalign{\bigskip}
&&\qquad\Bigl(\ldots,2^*2\,\tau\bigl\{k^{(R)}_{t_R}\bigr\}_{1\leq
t_R\leq s_R}(2^*2)^{n_R},\ldots,(2^*2)^{n_2}\tau\bigl\{k^{(1)}_{t_1}
\bigr\}_{1\leq t_1\leq s_1}(2^*2)^{n_1-m_1-1},(1^*1-0^*0),(2^*2)^{m_1}
i^*\Bigr)\cr
\noalign{\bigskip}
&&-\sum_{S=2}^{R}\sum_{m_{S}=0}^{n_{S}-1}(2-\delta_{m_{S},0}-
\delta_{m_{S},n_{S}-1})\,\Bigl(\sum_{l=1}^{S-1}\bigl(k^{(l)}_1
+k^{(l)}_2+\ldots+k^{(l)}_{s_l}+s_l+n_l-1)+m_{S}+1\Bigr)
\cdot\cr
\noalign{\bigskip}
&&\qquad\Bigl(\ldots,2^*2\,\tau\bigl\{k^{(R)}_{t_R}\bigr\}_{1\leq t_R\leq
s_R}(2^*2)^{n_R},\ldots,(2^*2)^{n_{S+1}}\tau\bigl\{
k^{(S)}_{t_{S}}\bigr\}_{1\leq t_{S}\leq s_{S}}
(2^*2)^{n_{S}-m_{S}-1},(1^*1-0^*0),\cr
\noalign{\bigskip}
&&\qquad\qquad\qquad\qquad
(2^*2)^{m_{S}}
\tau\bigl\{k^{(S-1)}_{t_{S-1}}\bigr\}_{1\leq t_{S-1}\leq s_{S-1}}(2^*2)^{
n_{S-1}},\ldots,(2^*2)^{n_2}\tau\bigl\{k^{(1)}_{t_1}\bigr\}_{1\leq t_1\leq
s_1}(2^*2)^{n_1}i^*\Bigr)\cr
\noalign{\bigskip}
&&-\sum_{m=0}^{\infty}(2-\delta_{m,0})\Bigl(\sum_{l=1}^R\bigl(k^{(l)}_1+
k^{(l)}_2+\ldots
+k^{(l)}_{s_l}+s_l+n_l-1\bigr)+m+1\Bigr)\cdot\cr
\noalign{\bigskip}
&&\qquad\Bigl(\ldots,2^*2,
(1^*1-0^*0),(2^*2)^{m}\,\tau\bigl\{k^{(R)}_{t_R}\bigr\}_{1\leq t_R\leq
s_R}(2^*2)^{n_R}
,\ldots,(2^*2)^{n_2}\tau\bigl\{k^{(1)}_{t_1}
\bigr\}_{1\leq t_1\leq s_1}(2^*2)^{n_1},i^*\Bigr)
\label{eq:h3}
\eea
with
\bea
\alpha\bigl(\bigl\{k^{(l)}_{t_l}\bigr\}_{1\leq t_l\leq s_l,\,1\leq
l\leq R},\;\{n_l\}_{1\leq l\leq R}\bigr)&=&\sum_{l=1}^R\bigl(s_l^2+(2
s_l-1)k^{(l)}_{s_l}+(2s_l-3)k^{(l)}_{s_l-1}+\ldots+3k^{(l)}_2+k^{(l)}_1
\bigr)\cr
\noalign{\bigskip}
&+&2\sum_{l=2}^Rs_l\Bigl(\sum_{m=1}^{l-1}\bigl(k^{(m)}_1+k^{(m)}_2
+\ldots+k^{(m)}_{s_m}+s_m+n_{m+1}-1\bigr)\Bigr)\cr
\noalign{\bigskip}
&-&\delta_{n_1,0}\,\delta_{i,0}-\delta_{n_1,0}\,\delta_{i,1}+2n_1
\sum_{l=1}^Rs_l
\label{eq:alpha}
\eea
The fourth and fifth line of (\ref{eq:h3}) give no contribution for
$n_1=0$. The configuration in the first term on the rhs of (\ref{eq:h3})
coincides with the configuration on the lhs. Obviously, all other
configurations on the rhs can be rewritten  as configurations of the
type (\ref{eq:conf1}) built from $r+1$ subsequences $1^*1-0^*0$ 
between subsequences $2^*2$ and the right border $\ldots,i^*\bigr)$.
Hence, for each $i=0,1,2$ the set of all configurations of the form
(\ref{eq:conf1}) with $r=0,1,2,\ldots$ is closed under the action of $H_0$.
For a given $i$,
the corresponding matrix elements of $H_0$ form a triangular matrix with
the diagonal matrix elements given by (\ref{eq:alpha}).
Since these are all smaller than zero for a configuration different
from $\bigl(\ldots,2^*2,2^*2,i^*\bigr)$, an eigenvector of $H_0$ with
eigenvalue zero is given by a linear combination of all configurations
(\ref{eq:conf1}) for each fixed $i$. An eigenvector of $H_0$
with eigenvalue $-2(n+1)$
is provided by a linear combination involving $\bigl(\ldots,2^*2,2^*2,
(1^*1-0^*0),(2^*2)^n,2^*\bigr)$ and all configurations generated from these
by repeated action of $H_0$. Among the latter, the configurations
$\bigl(\ldots,2^*2,2^*2,(1^*1-0^*0)^{1+2m},(2^*2)^{n-m},2^*\bigr)$ with
$1\leq m\leq n$ have the same value of the diagonal element 
(\ref{eq:alpha}) as $\bigl(\ldots,2^*2,2^*2,(1^*1-0^*0),(2^*2)^n,2^*
\bigr)$. It is easily verified that these don't occur in the appropriate
linear combination. Analogously, an eigenvector of $H_0$ can be
constructed as a linear combination of any configuration (\ref{eq:conf1})
and all configurations obtained from it by repeated action of $H_0$
with a different diagonal element (\ref{eq:alpha}). The corresponding
eigenvalue coincides with the diagonal element (\ref{eq:alpha}) of the
particular configuration chosen. In this context, diagonalizability is
a consequence of the particular form of the corner transfer matrix
Hamiltonian (\ref{eq:hctm1}), (\ref{eq:hlimdef}) and does not apply
to the $q\to0$-limit of the transfer matrix Hamiltonian acting on the
configurations (\ref{eq:conf1}), for example.

A similar structure applies to configurations $\bigl(\ldots,2^*2,2^*2,2^*
(0,2^*)^l(2,2^*)^n\bigr)$ with $n=0,1,2,\ldots$ and $l=1,2,3,\ldots$.
Given fixed values of $n,l$, the collection of states
\bea&&\bigl(\ldots,2^*2,2^*2,2^*(0,2^*)^l(2,2^*)^n\bigr)\cr
\noalign{\bigskip}
&&\bigl(\ldots,2^*2,2^*2,2^*0,(1^*1-0^*0),2^*(0,2^*)^{l-1}(2,2^*)^n\bigr)\cr
\noalign{\bigskip}
&&\bigl(\ldots,2^*2,2^*2,2^*(0,2^*)^{l-1}0,(1^*1-0^*0),2^*(2,2^*)^{n-1}\bigr)
\qquad\mbox{for}\;\;n>0\cr
\noalign{\bigskip}
&&\bigl(\ldots,2^*2,2^*2,2^*0,(1^*1-0^*0),2^*(0,2^*)^{l-2}0,
(1^*1-0^*0),2^*(2,2^*)^{n-1}\bigr)\qquad\mbox{for} \;\;n>0,l>1
\label{eq:conf1a}
\eea
and all configurations obtained from these by replacing one or more
subsequences $2^*2$ by $1^*1-0^*0$ is closed under the action of $H_0$.
Application of $H_0$ on one of these configurations with $r$ subsequences
$1^*1-0^*0$ yields the same configuration with a prefactor depending
on the position of these subsequences as well as further configurations,
each of them containing $r+1$ subsequences $1^*1-0^*0$. An eigenvector
of $H_0$ with an eigenvalue $\alpha$ given by the diagonal element of
$H_0$ on any configuration (\ref{eq:conf1a}) is given by a linear
combination of this configuration and all others generated from it by
repeated action of $H_0$ with the diagonal element of $H_0$ taking a value
different from $\alpha$.
Completely analogous statements hold true for the configurations
\bea
&&\bigl(\ldots,2^*2,2^*2,2^*0,0^*(2,2^*)^n\bigr)\cr
\noalign{\bigskip}
&&\bigl(\ldots,2^*2,2^*2,2^*(0,2^*)^{r_R}(2,2^*)^{n_R}(0,2^*)^{r_{R-1}}
(2,2^*)^{n_{R-1}}\ldots(0,2^*)^{r_1}(2,2^*)^{n_1}(0,0^*)^k(2,2^*)^{n_0}
\bigr)\cr
\noalign{\bigskip}
&&\bigl(\ldots,2^*2,2^*2,2^*(0,0^*)^k(2,2^*)^{n_0}(2,0^*)^{r_{R}}
(2,2^*)^{n_R}(2,0^*)^{r_{R-1}}(2,2^*)^{n_{R-1}}\ldots(2,0^*)^{r_1}(2,2^*
)^{n_1}\bigr)\cr
\noalign{\bigskip}
&&\bigl(\ldots,2^*2,2^*2,2^*(0,2^*)^{r_R}(2,2^*)^{n_R}(0,2^*)^{r_{R-1}}
(2,2^*)^{n_{R-1}}\ldots\cr
\noalign{\bigskip}
&&\qquad\ldots(0,2^*)^{r_1}(2,2^*)^{n_1}(0,0^*)^k(2,2^*)^{n_0}
(2,0^*)^{k_S}(2,2^*)^{m_S}(2,0^*)^{k_{S-1}}(2,2^*)^{m_{S-1}}\ldots(2,0^*)^{
k_1}(2,2^*)^{m_1}\bigr)
\eea
with $R,S>0$; $k=0,1$; $n,n_0,n_1,m_1\geq0$; $n_l,m_l>0$
for $l>1$ and $r_l,k_l>0\;\forall l$.

So fare, (\ref{eq:lim3}), (\ref{eq:lim5}) and the second line of
equations (\ref{eq:lim2}) do not enter the evaluation of $H_0$. A
simple example requiring (\ref{eq:lim3}) in addition is given by
\bea
&&H_0\Bigl(\bigl(\ldots,2^*2,2^*2,2^*1,1^*(2,2^*)^n\bigr)\Bigr)=
-(2n+1)\bigl(\ldots,2^*2,2^*2,2^*1,1^*(2,2^*)^n\bigr)\cr
\noalign{\bigskip}
&&\qquad+n\bigl(\ldots,2^*2,2^*2,(1^*1-0^*0),(2^*2)^{n-1},2^*\bigr)+(n+1)
\bigl(\ldots,2^*2,2^*2,(1^*1-0^*0),(2^*2)^n,2^*\bigr)\cr
\noalign{\bigskip}
&&\qquad-n\bigl(\ldots,2^*2,2^*2,2^*1,(1^*1-0^*0),1^*(2,2^*)^{n-1}\bigr)
-(n+1)\bigl(\ldots,2^*2,2^*2,2^*1,(1^*1-0^*0),1^*(2,2^*)^n\bigr)\cr
\noalign{\bigskip}
&&\qquad-2\sum_{m=0}^{n-2}(m+1)\,
\bigl(\ldots,2^*2,2^*2,2^*1,1^*2,(2^*2)^{n-m-2},
(1^*1-0^*0),(2^*2)^m,2^*\bigr)\cr
\noalign{\bigskip}
&&\qquad-2\sum_{m=0}^{\infty}
(n+m+2)\,\bigl(\ldots,2^*2,2^*2,(1^*1-0^*0),
(2^*2)^m,2^*1,1^*(2,2^*)^n\bigr)
\label{eq:h4}
\eea
The second line shows
that configurations of the type (\ref{eq:conf1}) with $i=2$ are generated by
repeated action of $H_0$ on $\bigl(\ldots,2^*2,2^*2,2^*1,1^*(2,2^*)^n
\bigr)$. As the following example reveals, application of $H_0$ on
configurations of the rhs of (\ref{eq:h4}) may produce several
configurations with one subsequence $1^*1-0^*0$ in addition to
those with two such subsequences:
\bea
&&H_0\Bigl(\bigl(\ldots,2^*2,2^*1,(1^*1-0^*0),1^*(2,2^*)^n\bigr)\Bigr)=
-2(n+1)\bigl(\ldots,2^*2,2^*1,(1^*1-0^*0),1^*(2,2^*)^n\bigr)\cr
\noalign{\bigskip}
&&\qquad+n\bigl(\ldots,2^*2,2^*2,2^*1,1^*2,(1^*1-0^*0),2^*(2,2^*)^{n-1}
\bigr)+(n+2)\bigl(\ldots,2^*2,2^*2,(1^*1-0^*0),2^*1,1^*(2,2^*)^{n}
\bigr)\cr
\noalign{\bigskip}
&&\qquad-n\bigl(\ldots,2^*2,2^*2,2^*1,(1^*1-0^*0)^2,1^*(2,2^*)^{n-1}\bigr)
-(n+2)\bigl(\ldots,2^*2,2^*2,2^*1,(1^*1-0^*0)^2,1^*(2,2^*)^{n}\bigr)\cr
\noalign{\bigskip}
&&\qquad-2\sum_{m=0}^{n-2}(m+1)\,
\bigl(\ldots,2^*2,2^*2,2^*1,(1^*1-0^*0),1^*2,
(2^*2)^{n-m-2},(1^*1-0^*0),(2^*2)^m,2^*\bigr)\cr
\noalign{\bigskip}
&&\qquad-2\sum_{m=0}^{\infty}(n+m+3)\,\bigl(\ldots,2^*2,
2^*2,(1^*1-0^*0),(2^*2)^m,2^*1,(1^*1-0^*0),1^*(2,2^*)^n\bigr)
\label{eq:h5}
\eea
In (\ref{eq:h4}) and (\ref{eq:h5}), the fourth line is not present for
$n=0,1$. Similar shifts of pieces $1^*1-0^*0$ occur for configurations 
arising from the action of $H_0$ on $\bigl(\ldots,2^*2,2^*2,2^*1,i^*2,
(2^*2)^n,2^*\bigr)$ and $\bigl(\ldots,2^*2,2^*2,1^*i,(2^*2)^n,2^*\bigr)$
with $i=0,2$, for example. They may be summarized by
\bea
&&H_0\Bigl(\bigl(\ldots
i,(1^*1-0^*0)^l,j^*i_n,j^*_ni_{n-1},\ldots i_2,j^*_2i_1,j^*_1\bigr)
\Bigr)=\cr
\noalign{\bigskip}
&&\qquad\qquad
\ldots+\delta_{i,1}(n+l+1)\bigl(\ldots2,(1^*1-0^*0),2^*1,
(1^*1-0^*0)^{l-1},j^*i_n,j^*_ni_{n-1},\ldots i_2,j^*_2i_1,j^*_1\bigr)+\ldots\cr
\noalign{\bigskip}
&&H_0\Bigl(\bigl(\ldots j,(1^*1-0^*0)^l,i^*2,2^*
i_n,j^*_ni_{n-1},\ldots i_2,j^*_2i_1,j^*_1\bigr)\Bigr)=\cr
\noalign{\bigskip}
&&\qquad\qquad\ldots+\delta_{i,1}(n+1)
\bigl(\ldots j,(1^*1-0^*0)^{l-1},1^*2,(1^*1-0^*0),2^*
i_n,j^*_ni_{n-1},\ldots i_2,j^*_2i_1,j^*_1\bigr)+\ldots
\label{eq:jump}
\eea
The contribution on the rhs of (\ref{eq:jump})
applies to all $i,i_1,j_1,i_2,j_2,\ldots,i_N,j_N=0,1,2$. 
Left of the subsequences indicated here, any sequence
with almost all entries equal to $2$ or $2^*$ may be inserted.
The limit (\ref{eq:lim3}) also enters the analysis of the set of
configurations 
$\bigl(\ldots,2^*2,2^*2,2^*i_1,j^*_1i_2,j^*_2(2,2^*)^n\bigr),\;n=0,1,2,
\ldots$, with the cases
$i_1=j_1=2$, $i_2=j_2=2$ and $j_1=i_2=0,1$ excluded:
\bea
&&H_0\Bigl(\bigl(\ldots,2^*2,2^*2,2^*i_1,j^*_1i_2,j^*_2(2,2^*)^n\bigr)\Bigr)
=\cr
\noalign{\bigskip}
&&-\Bigl((n+2)\bigl(1+y_{j_1,i_1,2}\bigr)+(n+1)\bigl(
y_{i_1,j_1,i_2}+y_{j_2,i_2,j_1}\bigr)+n\bigl(1+y_{i_2,j_2,2}
\bigr)\Bigr)\cdot
\bigl(\ldots,2^*2,2^*2,2^*i_1,j^*_1i_2,j^*_2(2,2^*)^n\bigr)
\cr
\noalign{\bigskip}
&&\qquad+\delta_{i_1,1}\delta_{j_1,1}(n+2)\,\bigl(\ldots,2^*2,2^*2,
(1^*1-0^*0),2^*i_2,j^*_2(2,2^*)^n\bigr)\cr
\noalign{\bigskip}
&&\qquad+\delta_{i_2,1}\delta_{j_2,1}\,n\,\bigl(\ldots,2^*2,2^*2,2^*i_1,
j^*_12,(1^*1-0^*0),(2^*2)^{n-1},2^*\bigr)\cr
\noalign{\bigskip}
&&\qquad-(n+2)\,\bigl(\ldots,2^*2,2^*2,2^*i_1,
(1^*1-0^*0),j_1^*i_2,j^*_2(2,2^*)^n\bigr)\cr
\noalign{\bigskip}
&&\qquad-n\,\bigl(\ldots,2^*2,2^*2,2^*i_1,
j^*_1i_2,(1^*1-0^*0),j_2^*(2,2^*)^{n-1}\bigr)\cr
\noalign{\bigskip}
&&\qquad-2\sum_{m=0}^{n-2}(m+1)\bigl(\ldots,2^*2,2^*2,2^*i_1,
j^*_1i_2,j^*_22,(2^*2)^{n-m-2},(1^*1-0^*0),(2^*2)^m,2^*\bigr)\cr
\noalign{\bigskip}
&&\qquad-2\sum_{m=0}^{\infty}(n+m+3)\,\bigl(\ldots,2^*2,2^*2,
(1^*1-0^*0),(2^*2)^m,2^*i_1,j^*_1i_2,j^*_2(2,2^*)^n\bigr)
\label{eq:h6}
\eea
For none of the configurations considered so fare, the (repeated)
action of $H_0$ creates configurations with a subsequence $1^*1+0^*0$.
Hence, (\ref{eq:lim1a}), (\ref{eq:lim1b}),
(\ref{eq:lim3}), (\ref{eq:lim4}) and the first
of equations (\ref{eq:lim2}) are sufficient
for its analysis. The following example 
involving also (\ref{eq:lim5}) and the remainder of (\ref{eq:lim2})
completes the evaluation of $(\ldots,2^*2,2^*2,2^*i_1,j^*_1i_2,j^*_2(2,
2^*)^n\bigr)$: 
\bea
&&H_0\Bigl(\bigl(\ldots,2^*2,2^*2,2^*i,(1^*1+0^*0),j^*(2,2^*)^n\bigl)\Bigl)
=\cr
\noalign{\bigskip}
&&-\bigl(2n+2-\delta_{i,2}+\delta_{j,2})\bigr)\cdot
\bigl(\ldots,2^*2,2^*2,2^*i,(1^*1+0^*0),j^*(2,2^*)^n\bigl)\cr
\noalign{\bigskip}
&&\qquad+\delta_{i,1}(n+2)\,\bigl(\ldots,2^*2,2^*2,(1^*1-0^*0),2^*1,j^*(
2,2^*)^n\bigr)\cr
\noalign{\bigskip}
&&\qquad+2(\delta_{i,1}+\delta_{j,1})(n+1)\bigl(\ldots,2^*2,2^*2,2^*i,
(1^*1-0^*0),j^*(2,2^*)^n\bigr)\cr
\noalign{\bigskip}
&&
\qquad+2(1-\delta_{i,1})(n+1)\,\bigl(\ldots,2^*2,2^*2,2^*1,1^*i,j^*
(2,2^*)^n\bigr)\cr
\noalign{\bigskip}
&&\qquad+2(1-\delta_{j,1})(n+1)\,\bigl(\ldots,2^*2,2^*2,2^*i,j^*1,1^*
(2,2^*)^n\bigr)\cr
\noalign{\bigskip}
&&\qquad+2(n+1)\,\bigl(\ldots,2^*2,2^*2,2^*i,
j^*(2,2^*)^n\bigr)\cr
\noalign{\bigskip}
&&\qquad+2(n+1)\,\bigl(\ldots,2^*2,2^*2,2^*i,
j^*(2,2^*)^{n+1}\bigr)\cr
\noalign{\bigskip}
&&\qquad+\delta_{j,1}n\,\bigl(\ldots,2^*2,2^*2,2^*i,1^*2,(1^*1-0^*0),
(2^*2)^{n-1},2^*\bigr)\cr
\noalign{\bigskip}
&&\qquad-(n+2)\,\bigl(\ldots,2^*2,2^*2,2^*i,(1^*1-0^*0),(1^*1+0^*0),
j^*(2,2^*)^n\bigr)\cr
\noalign{\bigskip}
&&\qquad-n\,\bigl(\ldots,2^*2,2^*2,2^*i,(1^*1+0^*0),(1^*1-0^*0),1^*(2,
2^*)^{n-1}\bigr)\cr
\noalign{\bigskip}
&&\qquad-2\sum_{m=0}^{n-2}(m+1)\,\bigl(\ldots,2^*2,2^*2,2^*i,(1^*1+
0^*0),j^*2,(2^*2)^{n-m-2},(1^*1-0^*0),(2^*2)^m,2^*\bigr)\cr
\noalign{\bigskip}
&&\qquad-2\sum_{m=0}^{\infty}(n+m+3)\,\bigl(\ldots,2^*2,2^*2,(1^*1-0^*0
),(2^*2)^m,2^*i,(1^*1+0^*0),j^*(2,2^*)^n\bigr)
\label{eq:h7}
\eea
In the fifth and sixth line of (\ref{eq:h7}), configurations included
in (\ref{eq:h6}) appear. The configuration in the fourth line is found in
$H_0\Bigl(\bigl(\ldots,2^*2,2^*2,2^*i,j^*(2,2^*)^n\bigr)\Bigr)$ which also
appears in the seventh line of (\ref{eq:h7}). As in the previous examples,
the configuration on the lhs of (\ref{eq:h7}) is not generated by
(repeated) action of $H_0$ on any of the other configurations found in
the rhs of the same equation. Further investigation
along these lines suggests to formulate this observation more generally. 
This is conveniently done by means of the notation
\bea
H_0\Bigl(\bigl(\ldots i_3,j^*_3i_2,j^*_2i_1,j^*_1\bigr)\Bigr)
=\sum_{i'_l,j'_l=
0,1,2}h_{\ldots i_3,\;j^*_3i_2,\;j^*_2i_1,\;j^*_1;\;\ldots i'_3,j'^*_3
i'_2,j'^*_2i'_1,j'^*_1}\cdot\bigl(\ldots i'_3,j'^*_3
i'_2,j'^*_2i'_1,j'^*_1\bigr)
\label{eq:hwrite}
\eea
Only finitely many values of $i_l,i'_l,j_l,j'_l$ differ from $2$
according to the boundary condition adopted here.
The following statement applies to all configurations
$\bigl(\ldots i_3,j^*_3i_2,j^*_2i_1,j^*_1\bigr)$ considered above:
\bea
&&h_{\ldots i_3,\;j^*_3i_2,\;j^*_2i_1,\;j^*_1;\;\ldots i'_3,j'^*_3
i'_2,j'^*_2i'_1,j'^*_1}\neq0\;\Longrightarrow\;\cr
\noalign{\bigskip}
&&\qquad\qquad\qquad\qquad
h_{\ldots i'_3,\;j'^*_3i'_2,\;j'^*_2i'_1,\;j'^*_1;\;\ldots i_3,j^*_3
i_2,j^*_2i_1,j^*_1}=\beta(\ldots i_3,j^*_3i_2,j^*_2i_1,j^*_1)\cdot
\prod_{l=1}^{\infty}\delta_{i_l,i'_l}\delta_{j_l,j'_l}
\label{eq:tri}
\eea
with
\bea
\beta(\ldots i_3,j^*_3i_2,j^*_2i_1,j^*_1)=-\sum_{l=1}^{\infty}
\bigl(y_{i_{l+1},j_{l+1},i_l}+y_{j_l,i_l,j_{l+1}}\bigr)
\label{eq:beta}
\eea
and $y_{i,j,k}$ given by (\ref{eq:lim1b}).
The function $\beta(\ldots i_3,j^*_3i_2,j^*_2i_1,j^*_1)$ is well defined
due to (\ref{eq:lim1b}) and the restriction $i_l=j_l=2$ for almost all $l$.
As the analysis reveals, (\ref{eq:tri}) remains valid for a large set of
configurations including the collection of all
$\bigl(\ldots i_3,j^*_3i_2,j^*_2i_1,j^*_1\bigr)$ which do not contain any
of the subsequences 
\bea
\begin{array}{c}
(1^*1+0^*0),(1^*1-0^*0)\cr
\noalign{\bigskip}
(1^*1-0^*0),(1^*1+0^*0)\cr
\noalign{\bigskip}
0^*1,1^*0
\end{array}
\label{eq:ndsubs}
\eea
Moreover, the elements of $H_0$ with respect to this restricted set of
configurations can be arranged as a trigonal matrix.
This can be seen introducing a number $\Omega\bigl( 
(\ldots i_3,j^*_3i_2,j^*_2i_1,j^*_1)\bigr)\in {\bf Z}_{0,+}$
for each configuration $\bigl(\ldots i_3,j^*_3i_2,j^*_2i_1,j^*_1\bigr)$:
\bea
&&\Omega\bigl((\ldots i_3,j^*_3i_2,j^*_2i_1,j^*_1)\bigr)\equiv\cr
\noalign{\bigskip}
&&\Omega^-\bigl((\ldots i_3,j^*_3i_2,j^*_2i_1,j^*_1)\bigr)+
\tilde{\Omega}^-\bigl((\ldots i_3,j^*_3i_2,j^*_2i_1,j^*_1)\bigr)
-\Omega^+\bigl((\ldots i_3,j^*_3i_2,j^*_2i_1,j^*_1)\bigr)
-\tilde{\Omega}^+\bigl((\ldots i_3,j^*_3i_2,j^*_2i_1,j^*_1)\bigr)\cr
\noalign{\bigskip}
&&\qquad-\omega^-\bigl((\ldots i_3,j^*_3i_2,j^*_2i_1,j^*_1)\bigr)
-\omega^+\bigl((\ldots i_3,j^*_3i_2,j^*_2i_1,j^*_1)\bigr)
\label{eq:ome}
\eea
Here $\Omega^-\bigl((\ldots i_3,j^*_3i_2,j^*_2i_1,j^*_1)\bigr)$ 
$\Bigl(\Omega^+\bigl((\ldots i_3,j^*_3i_2,j^*_2i_1,j^*_1)
\bigr)\Bigr)$ denotes the number of subsequences $1^*1-0^*0$ $\Bigl(1^*1+
0^*0\Bigr)$ found in the configuration $\bigl(\ldots i_3,j^*_3i_2,
j^*_2i_1,j^*_1\bigr)$ and ${1\over2}\Bigl(\tilde{\Omega}^+\bigl((
\ldots i_3,j^*_3i_2,j^*_2i_1,j^*_1\bigr)\Bigr)+\tilde{\Omega}^-\Bigl(
\bigl(\ldots i_3,j^*_3i_2,j^*_2i_1,j^*_1)\bigr)\Bigr)$ and
${1\over2}\Bigl(\tilde{\Omega}^+\bigl((
\ldots i_3,j^*_3i_2,j^*_2i_1,j^*_1)\bigr)-\tilde{\Omega}^-\bigl(
(\ldots i_3,j^*_3i_2,j^*_2i_1,j^*_1)\bigr)\Bigr)$
count the subsequences $1^*2,2^*1$ and $0^*2,2^*0$, respectively.
To each subsequence
\bea
&&j^*_1i_1,(1^*1+0^*0)^{m_1},j^{*(1)}_11,j^{*(1)}_21,\ldots,j^{*(1)}_{t_1}
1,(1^*1+0^*0)^{m_2},j^{*(2)}_11,j^{*(2)}_21,\ldots,j^{*(2)}_{t_2}1,
(1^*1+0^*0)^{m_3},\ldots\cr
&&\qquad\qquad\qquad\qquad\qquad\qquad\qquad\qquad\qquad
\ldots,(1^*1+0^*0)^{m_{S}},j^{*(s)}_11,j^{*(s)}_21,\ldots,j^{*(s)}_{t_S}
1,(1^*1+0^*0)^{m_{S+1}},j^*_2i_2
\label{eq:seq1}
\eea
with $S\geq1$; $m_{S+1}\geq0$; $m_l\geq1$ for $l\leq S$; $j^{*(l)}_{m_l}=
0,2$ for $1\leq m_l\leq t_l, \;1\leq l\leq S$ and the cases $i_m=j_m=0,1$
and $i_m=1,j_m=0,2$ excluded for $m=1,2$, the number 
\bea
\sum_{l=1}^Sm_l\sum_{l'=l}^St_{l'}
\label{eq:num1}
\eea
is assigned. Furthermore, to each subsequence
\bea
&&i^*_2j_2,(1^*1+0^*0)^{n_{R+1}},1^*i^{(R)}_{s_R},\ldots,1^*i^{(R)}_2,1^*
i^{(R)}_1,(1^*1+0^*0)^{n_R},\ldots\cr
\noalign{\bigskip}
&&\qquad
\ldots,(1^*1+0^*0)^{n_3},1^*i^{(2)}_{s_2},\ldots,1^*i^{(2)}_2,1^*
i^{(2)}_1,(1^*1+0^*0)^{n_2},1^*i^{(1)}_{s_1},\ldots,1^*i^{(1)}_2,1^*
i^{(1)}_1,(1^*1+0^*0)^{n_1},i^*_1j_1
\label{eq:seq2}
\eea
with $R\geq1$; $n_{R+1}\geq0$; $n_l\geq1$ for $l\leq R$; $i^{(l)}_{m_l}=
0,2$ for $1\leq m_l\leq s_l,\;1\leq l\leq R$ and the above restrictions
on $i_m,j_m,\;m=1,2$, the number
\bea
\sum_{l=1}^Rn_l\sum_{l'=l}^Rs_{l'}
\label{eq:num2}
\eea
is attributed. $\omega^+\bigl((\ldots i_3,j^*_3i_2,j^*_2i_1,
j^*_1)\bigr)$ equals the sum of numbers (\ref{eq:num1}) and
(\ref{eq:num2}) found for the configuration $(\ldots i_3,j^*_3i_2,j^*_2i_1,
j^*_1))$. Similarly, the numbers
\bea
\sum_{l=2}^{S+1}m_l\sum_{l'=1}^{l-1}t_{l'}\qquad\mbox{and}\qquad
\sum_{l=2}^{R+1}n_l\sum_{l'=1}^{l-1}s_{l'}
\label{eq:num3}
\eea
are assigned to each subsequence obtained from (\ref{eq:seq1})
and (\ref{eq:seq2}) by replacing each term $1^*1+0^*0$
by $1^*1-0^*0$. 
Then $\omega^-\bigl((\ldots i_3,j^*_3i_2,j^*_2i_1,j^*_1)\bigr)$
denotes the sum of all numbers (\ref{eq:num3}) related to such subsequences
present in the configuration $(\ldots i_3,j^*_3i_2,j^*_2i_1,j^*_1))$.
Direct examination of the limits (\ref{eq:lim2})-(\ref{eq:lim5}) reveals
that two configurations $(\ldots i_3,j^*_3i_2,j^*_2i_1,j^*_1)$ and
$(\ldots i'_3,j'^*_3i'_2,j'^*_2i'_1,j'^*_1)$ with
$h_{\ldots i_3,\;j^*_3i_2,\;j^*_2i_1,\;j^*_1;\;\ldots i'_3,j'^*_3
i'_2,j'^*_2i'_1,j'^*_1}\neq0$ and $(\ldots i_3,j^*_3i_2,j^*_2i_1,j^*_1)
\neq(\ldots i'_3,j'^*_3i'_2,j'^*_2i'_1,j'^*_1)$ satisfy
\bea
\Omega\bigl((\ldots i'_3,j'^*_3i'_2,j'^*_2i'_1,j'^*_1)\bigr)>
\Omega\bigl((\ldots i_3,j^*_3i_2,j^*_2i_1,j^*_1)\bigr)
\label{eq:prop1}
\eea
if the configuration $(\ldots i'_3,j'^*_3i'_2,j'^*_2i'_1,j'^*_1)$
is not obtained from $(\ldots i_3,j^*_3i_2,j^*_2i_1,j^*_1)$ either
by substituting one subsequence $\bigl(1^*1+(-1)^k0^*0\bigr)\bigl(1^*1-
(-1)^k0^*0\bigr)$ by $\bigl(1^*1-(-1)^k0^*0\bigr)\bigl(
1^*1+(-1)^k0^*0\bigr)$ or $0^*1,1^*0$, where $k=0,1$, 
or by replacing one subsequence $0^*1,1^*0$ by $(1^*1+0^*0)(1^*1-0^*0)$
or $(1^*1-0^*0)(1^*1+0^*0)$. In these cases, the limits (\ref{eq:lim2}),
(\ref{eq:lim3}) and (\ref{eq:lim5}) yield for
$i,j,i_l,j_l=0,1,2$:
\bea
&&H_0\Bigl(\bigl(\ldots,i,(1^*1+0^*0),(1^*1-0^*0),j^*i_n,j^*_ni_{n-1},
\ldots i_2,j^*_2i_1,j^*_1\bigr)\Bigr)=\cr
\noalign{\bigskip}
&&\qquad\qquad\qquad\qquad\ldots+(n+2)
\bigl(\ldots,i,(1^*1-0^*0),(1^*1+0^*0),j^*i_n,j^*_ni_{n-1},
\ldots i_2,j^*_2i_1,j^*_1\bigr)\cr
\noalign{\bigskip}
&&\qquad\qquad\qquad\qquad\qquad-2(n+2)\bigl(\ldots i,0^*1,1^*0,
j^*i_n,j^*_ni_{n-1},\ldots i_2,j^*_2i_1,j^*_1\bigr)+\ldots\cr
\noalign{\bigskip}
&&H_0\Bigl(\bigl(\ldots,i,(1^*1-0^*0),(1^*1+0^*0),j^*i_n,j^*_ni_{n-1},
\ldots i_2,j^*_2i_1,j^*_1\bigr)\Bigr)=\cr
\noalign{\bigskip}
&&\qquad\qquad\qquad\qquad\ldots+(n+1)
\bigl(\ldots,i,(1^*1+0^*0),(1^*1-0^*0),j^*i_n,j^*_ni_{n-1},
\ldots i_2,j^*_2i_1,j^*_1\bigr)\cr
\noalign{\bigskip}
&&\qquad\qquad\qquad\qquad\qquad-2(n+1)\bigl(\ldots i,0^*1,1^*0,
j^*i_n,j^*_ni_{n-1},\ldots i_2,j^*_2i_1,j^*_1\bigr)+\ldots
\label{eq:heff1}
\eea
In both cases, the first contribution stems from (\ref{eq:lim2}) and
the second from (\ref{eq:lim5}). Furthermore,
\bea
&&H_0\Bigl(2\bigl(\ldots i,0^*1,1^*0,j^*i_n,j^*_ni_{n-1},\ldots i_2,j^*_2
i_1,j^*_1\bigr)\Bigr)=\cr
\noalign{\bigskip}
&&\qquad\qquad\qquad\qquad\ldots+(n+2)
\bigl(\ldots,i,(1^*1-0^*0),(1^*1+0^*0),j^*i_n,j^*_ni_{n-1},
\ldots i_2,j^*_2i_1,j^*_1\bigr)\cr
\noalign{\bigskip}
&&\qquad\qquad\qquad\qquad\qquad+(n+1)
\bigl(\ldots,i,(1^*1+0^*0),(1^*1-0^*0),j^*i_n,j^*_ni_{n-1},
\ldots i_2,j^*_2i_1,j^*_1\bigr)+\ldots
\label{eq:heff2}
\eea
where both contributions arise from (\ref{eq:lim3}). As in (\ref{eq:jump}),
left of the entry $i$ any sequence with almost all entries fixed by $2$
or $2^*$ may be chosen. 
Together with the limits (\ref{eq:lim1a}), (\ref{eq:lim1b}), the property
(\ref{eq:prop1}) implies triangularity of the matrix formed by the
elements of $H_0$ for the restricted set of configurations if arranged
in an order indicated
by increasing numbers $\Omega(\ldots i_3,j^*_3i_2,j^*_2i_1,j^*_1)$ .

\subsection{A particular set of subsequences}
\label{sec:nontrset}

This subsection specializes on a set of configurations for which the
matrix elements of $H_0$ don't show a triangular form in the basis
(\ref{eq:state}).
As apparent from (\ref{eq:heff1}),
(\ref{eq:heff2}), $H_0$ couples configurations which differ only by
maximal subsequences consisting
of $K-L$ terms $(1^*1+0^*0)$, $M-L$ terms $(1^*1-0^*0)$ and $L$ terms
$=0^*1,1^*0$ for fixed pairs $K,M\geq1$ and $0\leq L\leq \min(K,M)$.
Explicitly, these configurations are written
\begin{eqnarray*}
\bigl(\ldots,j^*_{n+3}i_{n+2},j^*_{n+2}i_{n+1},(0^*1,1^*0)^{n_1},j^*_{n+1}
i_{n},j^*_{n}i_{n-1},\ldots,j^*_2i_1,j^*_1\bigr)
\end{eqnarray*}
with $n\geq0$ and $n_1\geq1$ or
\bea
&&\Bigl(\ldots,j^*_{n+3}i_{n+2},j^*_{n+2}i_{n+1},(0^*1,1^*0)^{n_0}\Bigl\{
\prod_{t_1=1}^{s_1}(1^*1+(-1)^{\sigma_{t_1}}0^*0)\Bigr\}\;
(0^*1,1^*0)^{n_1}
\Bigl\{\prod_{t_2
=1}^{s_2}(1^*1+(-)^{\sigma_{t_2}}0^*0)\Bigr\},\ldots\cr
\noalign{\bigskip}
&&\qquad\ldots,(0^*1,1^*0)^{n_{R-1}}\Bigl\{\prod_{t_R=1}^{s_R}(1^*1+
(-1)^{\sigma_{t_R}}0^*0)\Bigr\}\;(0^*1,1^*0)^{n_{R}},j^*_{n+1}i_{n},
j^*_{n}i_{n-1},\ldots,j^*_2i_1,j^*_1\Bigr)
\label{eq:seq3}
\eea
where $R\geq1$; $n,n_0,n_{R}\geq0$; $n_l\geq1$ for $0<l<R$;
$s_l\geq1$ and $\sigma_{t_l}=0,1$. In (\ref{eq:seq3}), the choices
$j_{n+2}=i_{n+1}=0,1$ or $j_{n+1}=i_{n}=0,1$ as well as $j_{n+3}=i_{n+1}=0$,
$i_{n+2}=j_{n+2}=1$ or $j_{n+1}=i_{n-1}=0$, $i_{n}=j_{n}=1$ are excluded.
Any diagonal element of $H_0$ for the
configurations (\ref{eq:seq3}) depends on $n$,
$n_l$ and $\sigma_{t_l}$ and on the remaining entries $i_l,j_l$.
The contribution of the latter is the same
for all diagonal elements and will be denoted by
$-\kappa$ below. As an example, equations
(\ref{eq:lim1a}) and(\ref{eq:lim1b}) yield $\kappa=2n+1+K+M$
for the configuration $\bigl(\ldots,2^*2,2^*2,
(1^*1+0^*0)^K,(1^*1-0^*0)^M,(2^*2)^n,2^*\bigr)$.
The collection of matrix elements of $H_0$ with respect to all
configurations (\ref{eq:seq3}) for a given set of values $n,\,\kappa$
and $K,M$ may be called $h^{(n,\kappa)}$.
For the configurations (\ref{eq:seq3}), the diagonal elements of
$h^{(n,\kappa)}$ are given by
\bea
-\kappa-n_{1}(2n_1+2n+1)
\label{eq:dia1}
\eea
for the first configuration in (\ref{eq:seq3}) and by
\bea
-\kappa-n_{R}(2n_R+2n+1)-\sum_{l=1}^Rn_{l-1}\Bigl(2n+1+2n_{l-1}
+2\sum_{l'=l}^R(2n_{l'}+s_{l'})\Bigr)
\label{eq:dia2}
\eea
for all others. All other nonvanishing elements of $h^{(n,\kappa)}$ follow
directly from (\ref{eq:heff1}) and (\ref{eq:heff2}).
Within the subsequences (\ref{eq:seq3}), the symbols $+,\,-$ and $\circ$
will be used below
to abbreviate the terms $1^*1+0^*0, 1^*1-0^*0$ and twice
the term $0^*1,1^*0$, respectively. Explicit reference to  the entries
$i_l,j_l$ in the configurations (\ref{eq:seq3}) will be omitted in most
of the remainder of this section. Then the dependence on $n$ and
$\{i_l,j_l\}$ will be reminded by a subscript $n,\kappa$.
Equations (\ref{eq:heff1}), (\ref{eq:heff2}) imply
\bea
h^{(n,\kappa)}_{\bigl(b\bigl\{\circ-(-+)\bigr\}a\bigr);\;
\bigl(b\bigl\{\circ+(-+)\bigr\}a\bigr)}&=&h^{(n,\kappa)}_{\bigl(b
\bigl\{\circ-(-+)\bigr\}a\bigr);\;\bigl(b+-a\bigr)}=0\cr
\noalign{\bigskip}
h^{(n,\kappa)}_{\bigl(b\bigl\{\circ-(+-)\bigr\}a\bigr);\;
\bigl(b\bigl\{\circ+(+-)\bigr\}a\bigr)}&=&h^{(n,\kappa)}_{\bigl(b
\bigl\{\circ-(+-)\bigr\}a\bigr);\;\bigl(b-+a\bigr)}=0
\label{eq:heffp1}
\eea
where $a,\,b$ abbreviate parts of a subsequence built from
$+,\,-$ and $\circ$, $\bigl(b\bigl\{\circ\pm(+-)
\bigr\}a\bigr)_{n,\kappa}=\bigl(b\circ a\bigr)_{n,\kappa}\pm\bigl(b
+-a\bigr)_{n,\kappa}$, $\bigl(b\bigl\{\circ\pm(-+)
\bigr\}a\bigr)_{n,\kappa}=\bigl(b\circ a\bigr)_{n,\kappa}\pm\bigl(b
-+a\bigr)_{n,\kappa}$ and the notation
$h^{(n,\kappa)}\bigl((c)_{n,\kappa}\bigr)=\sum_{c'}h^{(n,\kappa)}_{(c);\,(
c')}
(c')_{n,\kappa}$ with subsequences $c,\,c'$ composed from $+,-$ and
$\circ$ is used.
With (\ref{eq:heffp1}), (\ref{eq:heff1}), (\ref{eq:heff2}) and
(\ref{eq:dia1}), (\ref{eq:dia2}) it is easily verified that the
configurations
\begin{eqnarray*}
&&\Bigl((-)^{M-1}\Bigl\{\circ-\bigl(-+\bigr)\Bigr\}(+)^{K-1}\Bigr)_{n,
\kappa}
\end{eqnarray*}
and
\bea
&&\Bigl((-)^{r_0}\Bigl\{\circ-\bigl(-+\bigr)\Bigr\}
(+)^{s_0}\Bigl\{\circ-\bigl(+-\bigr)\Bigr\}(-)^{r_1}
\Bigl\{\circ-\bigl(-+\bigr)\Bigr\}(+)^{s_1}\Bigl\{\circ-\bigl(+-\bigr)
\Bigr\}(-)^{r_2}\ldots\cr
\noalign{\bigskip}
&&\qquad\qquad\qquad
\ldots(+)^{s_{m-1}}\Bigl\{\circ-\bigl(+-\bigr)\Bigr\}(-)^{r_m}\Bigl\{
\circ-\bigl(-+\bigr)\Bigr\}(+)^{s_m}\Bigr)_{n,\kappa}
\label{eq:eig1}
\eea
with $m=1,2,3,\ldots$; $r_l,s_l=0,1,2,\ldots$ and $r_0=M-2m-1-\sum_{
l=1}^mr_l\geq0$; $s_0=K-2m-1-\sum_{l=1}^ms_l\geq0$
are eigenvectors of $h^{(n,\kappa)}$ with eigenvalues
\bea
-(\kappa+n+K+1)\qquad\mbox{and}\qquad
-\Bigl(\kappa+4m^2+m(n+1)+(m+1)(n+2)+K-1+2\sum_{l=1}^ml(r_l+s_l)\Bigr).
\label{eq:eigv1}
\eea
Similarly, the configurations
\bea
&&\Bigl((-)^{\tilde r_0}\Bigl\{\circ-\bigl(-+\bigr)\Bigr\}
(+)^{\tilde s_0}\Bigl\{\circ-\bigl(+-\bigr)\Bigr\}(-)^{r_1}
\Bigl\{\circ-\bigl(-+\bigr)\Bigr\}(+)^{s_1}\Bigl\{\circ-\bigl(+-\bigr)
\Bigr\}(-)^{r_2}\ldots\cr
\noalign{\bigskip}
&&\qquad\qquad\qquad
\ldots(+)^{s_{m-1}}\Bigl\{\circ-\bigl(+-\bigr)\Bigr\}(-)^{r_m}\Bigr)_{n,
\kappa}
\label{eq:eig2}
\eea
with  $M-2m-\sum_{l=1}^mr_l=\tilde r_0\geq0$; $K-2m-\sum_{l=1}^{
m-1}s_l=\tilde s_0\geq0$ are eigenvectors with eigenvalues
\bea
-\Bigl(\kappa+2(m-1)(2m-1)+m(n+1)+m(n+2)+K+2\sum_{l=1}^mlr_l+2\sum_{l
=1}^{m-1}ls_l\Bigr)
\label{eq:eigv2}
\eea
Exchanging $+$ and $-$ in (\ref{eq:eig1}) and (\ref{eq:eig2}) as well as
$K$ and $M$ in the restrictions for $r_0,s_0$ and $\tilde r_0,\tilde s_0$
yields further eigenvectors. The eigenvalues for the eigenvectors
resulting from (\ref{eq:eig1}) follow from (\ref{eq:eigv1})
by the replacement $K\to M-1$ while those for the eigenvectors
obtained from (\ref{eq:eig2}) follow from (\ref{eq:eigv2}) by the
substitution $K\to M$.

Further analysis reveals that the matrix elements of $h^{(n,\kappa)}$
exhibit a triangular structure with respect to a suitable basis
of configurations composed from $+,\,-,\,\bigl\{\circ-(-+)
\bigr\},\;\bigl\{\circ-(+-)\bigr\}$ and $\circ$ such that
$+$ and $-$ are adjacent only as parts of the linear combinations
$\bigl\{\circ-(-+)\bigr\}$ and $\bigl\{\circ-(+-)\bigr\}$.
Such a basis configuration may be written
\bea
\Bigl((\circ)^{n_0}\,c_1\,(\circ)^{n_1}\,c_2\,(\circ)^{n_2}\,c_3\,(\circ
)^{n_3}\ldots
c_{R-1}\,(\circ)^{n_{R-1}}\,c_R\,(\circ)^{n_R}\Bigr)_{n,\kappa}
\label{eq:seq4}
\eea
where $n_0,n_R\geq0$; $n_l>0$ for $0<l<R$ and the parts $c_l$ contain
any sequence of $+,\,-,\,\bigl\{\circ-(-+)\bigr\}$ and $\bigl\{\circ-(+-)
\bigr\}$ provided that only the terms $+,\,\bigl\{\circ-(-+)\bigr\}$ or
$\bigl\{\circ-(+-)\bigr\}$ are neighbors of $+$ and only $-,
\,\bigl\{\circ-(-+)\bigr\}$ or $\bigl\{\circ-(+-)\bigr\}$ are next to $-$.
For an arbitrary part $a$ of $c_l$, the length $\lambda(a)$ counts once
each symbol $+$ or $-$ found in $a$ and twice each linear combination  
$\bigl\{\circ-(-+)\bigr\}$ or $\bigl\{\circ-(+-)\bigr\}$ there. To any
decomposition $c_l=\bigl(b\bigl\{\circ-\bigl(\rho,-\rho\bigr)\bigr\}
a\bigr)$ with $\rho=\pm$ the number $n+1+\delta_{\rho,-}+\lambda(a)$
is associated.
Summing these numbers for all such decompositions of $c_l$ yields a
number $\gamma(c_l)$. In terms of these notations, 
(\ref{eq:heff1}), (\ref{eq:heff2}) and (\ref{eq:dia1}), (\ref{eq:dia2})
yield the diagonal matrix elements referring to the basis configurations
(\ref{eq:seq4}):
\bea
h^{(n,\kappa)}_{\bigl((\circ)^{n_0}\,c_1\,(\circ)^{n_1}\,c_2\,(\circ)^{n_2}
\ldots c_R\,(\circ)^{n_R}\bigr);\;\bigl((\circ)^{n_0}\,c_1\,(\circ)^{n_1}
\,c_2\,(\circ)^{n_2}\ldots c_R\,(\circ)^{n_R}\bigr)}=-\kappa-\sum_{l=1}^R
\gamma(c_l)
\label{eq:dia3}
\eea
One type of nondiagonal elements of $h^{(n,\kappa)}$ in the basis
(\ref{eq:seq4}) amounts to exchanging 
one linear combination $\bigl\{\circ-(\rho,-\rho)\bigr\}$ 
with a neighboring a single symbol $+$ or $-$:
\bea
h^{(n,\kappa)}_{\bigl(b\bigl\{\circ-(-+)\bigr\}-a\bigr);\;
\bigl(b-\bigl\{\circ-(-+)\bigr\}a\bigr)}&=&n+2+\lambda(a)\qquad\qquad
\qquad b\neq(\ldots+\cr
\noalign{\bigskip}
h^{(n,\kappa)}_{\bigl(b+\bigl\{\circ-(-+)\bigr\}a\bigr);\;
\bigl(b\bigl\{\circ-(-+)\bigr\}+a\bigr)}&=&n+3+\lambda(a)\qquad\qquad
\qquad a\neq-\ldots)\cr
\noalign{\bigskip}
h^{(n,\kappa)}_{\bigl(b\bigl\{\circ-(+-)\bigr\}+a\bigr);\;
\bigl(b+\bigl\{\circ-(+-)\bigr\}a\bigr)}&=&n+1+\lambda(a)\qquad\qquad
\qquad b\neq(\ldots-\cr
\noalign{\bigskip}
h^{(n,\kappa)}_{\bigl(b-\bigl\{\circ-(+-)\bigr\}a\bigr);\;
\bigl(b\bigl\{\circ-(+-)\bigr\}-a\bigr)}&=&n+2+\lambda(a)\qquad\qquad
\qquad a\neq+\ldots)
\label{eq:jump1}
\eea
In the cases excluded in the right column of (\ref{eq:jump1}), the
following matrix elements are found:
\bea
h^{(n,\kappa)}_{\bigl(b+\bigl\{\circ-(-+)\bigr\}-a\bigr);\;\bigl(b
\circ\bigl\{\circ-(-+)\bigr\}a\bigr)}
=-h^{(n,\kappa)}_{\bigl(b+\bigl\{\circ-(-+)\bigr\}-a\bigr);\;\bigl(b
\bigl\{\circ-(+-)\bigr\}\bigl\{\circ-(-+)\bigr\}a\bigr)}
&=&n+2+\lambda(a)\cr
\noalign{\bigskip}
h^{(n,\kappa)}_{\bigl(b+\bigl\{\circ-(-+)\bigr\}-a\bigr);\;\bigl(b
\bigl\{\circ-(-+)\bigr\}\circ a\bigr)}
=-h^{(n,\kappa)}_{\bigl(b+\bigl\{\circ-(-+)\bigr\}-a\bigr);\;\bigl(b
\bigl\{\circ-(-+)\bigr\}\bigl\{\circ-(+-)\bigr\}a\bigr)}
&=&n+4+\lambda(a)\cr
\noalign{\bigskip}
h^{(n,\kappa)}_{\bigl(b-\bigl\{\circ-(+-)\bigr\}+a\bigr);\;\bigl(b
\circ\bigl\{\circ-(+-)\bigr\}a\bigr)}
=-h^{(n,\kappa)}_{\bigl(b-\bigl\{\circ-(+-)\bigr\}+a\bigr);\;\bigl(b
\bigl\{\circ-(-+)\bigr\}\bigl\{\circ-(+-)\bigr\}a\bigr)}
&=&n+1+\lambda(a)\cr
\noalign{\bigskip}
h^{(n,\kappa)}_{\bigl(b-\bigl\{\circ-(+-)\bigr\}+a\bigr);\;\bigl(b
\bigl\{\circ-(+-)\bigr\}\circ a\bigr)}
=-h^{(n,\kappa)}_{\bigl(b-\bigl\{\circ-(+-)\bigr\}+a\bigr);\;\bigl(b
\bigl\{\circ-(+-)\bigr\}\bigl\{\circ-(-+)\bigr\}a\bigr)}
&=&n+3+\lambda(a)
\label{eq:jump2}
\eea
A second type of nondiagonal matrix elements reduces higher powers
of $\bigl\{\circ-(\rho,-\rho)\bigr\}$ according to
\bea
h^{(n,\kappa)}_{\bigl(b\bigl\{\circ-(-+)\bigr\}^2a\bigr);\;
\bigl(b-\bigl\{\circ-(-+)\bigr\}+a\bigr)}&=&-\bigl(n+3+\lambda(a)\bigr)
\qquad\qquad b\neq(\ldots+,\;\;\;a\neq -\ldots)\cr
\noalign{\bigskip}
h^{(n,\kappa)}_{\bigl(b\bigl\{\circ-(+-)\bigr\}^2a\bigr);\;
\bigl(b+\bigl\{\circ-(+-)\bigr\}-a\bigr)}&=&-\bigl(n+2+\lambda(a)\bigr)
\qquad\qquad b\neq(\ldots-,\;\;\;a\neq +\ldots)
\label{eq:square1}
\eea
The corresponding matrix elements in the cases $a=-\ldots)$ or $b=(\ldots+$
are
\bea
h^{(n,\kappa)}_{\bigl(b+\bigl\{\circ-(-+)\bigr\}^2a\bigr);\;\bigl(b\circ
\bigl\{\circ-(-+)\bigr\}+a\bigr)}=-h^{(n,\kappa)}_{\bigl(b+\bigl\{
\circ-(-+)\bigr\}^2a\bigr);\;\bigl(b\bigl\{\circ-(+-)\bigr\}
\bigl\{\circ-(-+)\bigr\}+a\bigr)}&=&-\bigl(n+3+\lambda(a)\bigr)
\cr
\noalign{\bigskip}
h^{(n,\kappa)}_{\bigl(b\bigl\{\circ-(-+)\bigr\}^2-a\bigr);\;\bigl(b-
\bigl\{\circ-(-+)\bigr\}\circ a\bigr)}=-h^{(n,\kappa)}_{\bigl(b\bigl\{
\circ-(-+)\bigr\}^2-a\bigr);\;\bigl(b-
\bigl\{\circ-(-+)\bigr\}\bigl\{\circ-(+-)\bigr\}a\bigr)}&=&-\bigl(
n+4+\lambda(a)\bigr)
\label{eq:square2}
\eea
where $a\neq-\ldots)$ in the first and $b\neq (\ldots+$ in the second
line, and  
\bea
h^{(n,\kappa)}_{\bigl(b+\bigl\{\circ-(-+)\bigr\}^2-a\bigr);\;\bigl(b\circ
\bigl\{\circ-(-+)\bigr\}\circ a\bigr)}&=&-
h^{(n,\kappa)}_{\bigl(b+\bigl\{\circ-(-+)\bigr\}^2-a\bigr);\;\bigl(b\circ
\bigl\{\circ-(-+)\bigr\}\bigl\{\circ-(+-)\bigr\}a\bigr)}=\cr
\noalign{\bigskip}
=-h^{(n,\kappa)}_{\bigl(b+\bigl\{\circ-(-+)\bigr\}^2-a\bigr);\;\bigl(b
\bigl\{\circ-(+-)\bigr\}\bigl\{\circ-(-+)\bigr\}\circ a\bigr)}&=&
h^{(n,\kappa)}_{\bigl(b+\bigl\{\circ-(-+)\bigr\}^2-a\bigr);\;\bigl(b
\bigl\{\circ-(+-)\bigr\}\bigl\{\circ-(-+)\bigr\}\bigl\{\circ-(+-)\bigr\}
a\bigr)}=\cr
\noalign{\bigskip}
&&\qquad\qquad=-\bigl(n+4+\lambda(a)\bigr)
\label{eq:square3}
\eea
Exchanging $+$ and $-$ in (\ref{eq:square2}) and
(\ref{eq:square3}) and replacing $\lambda(a)$ by $\lambda(a)-1$ in the
rightmost parts of the equations covers the cases excluded in the 
second line of (\ref{eq:square1}).
Finally, a third type of nondiagonal matrix elements substitutes a single
symbol $\circ$ by the linear combinations $\bigl\{\circ-(-+)\bigr\}$
and $\bigl\{\circ-(+-)\bigr\}$:
\bea
h^{(n,\kappa)}_{\bigl(b\circ a\bigr);\;\bigl(b\bigl\{\circ-(-+)\bigr\}
a\bigr)}&=&-\bigl(n+2+\lambda(a)\bigr)\cr
\noalign{\bigskip}
h^{(n,\kappa)}_{\bigl(b\circ a\bigr);\;\bigl(b\bigl\{\circ-(+-)\bigr\}
a\bigr)}&=&-\bigl(n+1+\lambda(a)\bigr)
\label{eq:circre}
\eea
Equations (\ref{eq:jump1})-(\ref{eq:circre}) characterize
the action of $h^{(n,\kappa)}$ on any subsequence of the form
(\ref{eq:seq4}).
A particular set of subsequences built from $+,\,-,\,\bigl\{\circ-(-+)
\bigr\},\,\bigl\{\circ-(+-)\bigr\}$ and $\circ$ is generated by the action
of $h^{(n,\kappa)}$ on 
\begin{eqnarray*}
&&\Bigl((-)^{M-1}\Bigl\{\circ-\bigl(+-\bigr)\Bigr\}(+)^{K-1}\Bigr)_{n,
\kappa}
\end{eqnarray*}
and
\bea
&&\Bigl((-)^{r_0}\Bigl\{\circ-\bigl(+-\bigr)\Bigr\}
(+)^{s_0}\Bigl\{\circ-\bigl(-+\bigr)\Bigr\}(-)^{r_1}
\Bigl\{\circ-\bigl(+-\bigr)\Bigr\}(+)^{s_1}\Bigl\{\circ-\bigl(-+\bigr)
\Bigr\}(-)^{r_2}\ldots\cr
\noalign{\bigskip}
&&\qquad\qquad\qquad
\ldots(+)^{s_{m-1}}\Bigl\{\circ-\bigl(-+\bigr)\Bigr\}(-)^{r_m}\Bigl\{
\circ-\bigl(+-\bigr)\Bigr\}(+)^{s_m}\Bigr)_{n,\kappa}
\label{eq:start1}
\eea
with $m,r_m,r_l,s_l>0$ for $0<l<m;\;s_m\geq0$, $r_0=M-2m-1-\sum_{l=1}^mr_l
\geq0$, $s_0=K-2m-1-\sum_{l=1}^ms_l>0$ and
\bea
&&\Bigl((-)^{\tilde r_0}\Bigl\{\circ-\bigl(+-\bigr)\Bigr\}
(+)^{\tilde s_0}\Bigl\{\circ-\bigl(-+\bigr)\Bigr\}(-)^{r_1}
\Bigl\{\circ-\bigl(+-\bigr)\Bigr\}(+)^{s_1}\Bigl\{\circ-\bigl(-+\bigr)
\Bigr\}(-)^{r_2}\ldots\cr
\noalign{\bigskip}
&&\qquad\qquad\qquad
\ldots(+)^{s_{m-1}}\Bigl\{\circ-\bigl(-+\bigr)\Bigr\}(-)^{r_m}\Bigr)_{n,
\kappa}
\label{eq:start2}
\eea
with $m,r_l,s_l>0$ for $0<l<m;\;r_m\geq0$, $\tilde r_0=M-2m-\sum_{l=1}^m
r_l\geq0$ and $\tilde s_0=K-2m-\sum_{l=1}^{m-1}s_l>0$ as well as on the
subsequences obtained from (\ref{eq:start1}), (\ref{eq:start2})
via the replacements $+\leftrightarrow-$ and $K\leftrightarrow M$.
A subsequence of the type (\ref{eq:start1}) or (\ref{eq:start2}) does not
arise in the action of $h^{(n,\kappa)}$ on any other subsequence.
The subsequences created by repeated action of $h^{(n,\kappa)}$ on the
subsequences (\ref{eq:start1}), (\ref{eq:start2}) all have the form
(\ref{eq:seq4}) and may be classified
as follows. A first group consists of all subsequences (\ref{eq:seq4})
with $R=1$ and $n_0=n_1=0$. These are subsequences without a single
symbol $\circ$. The second group collects all configurations with one
single $\circ$ contained in one of the parts
\bea
\bigl(-\rho_1\bigr)^{u-u_1}\Bigl\{\circ-\bigl(\rho_1,-\rho_1\bigr)\Bigr\}
\bigl(-\rho_1\bigr)^{u_1}&\circ&\bigl(-\rho_2\bigr)^{u'_1}\Bigl\{\circ-
\bigl(-\rho_2,\rho_2\bigr)\Bigr\}\bigl(-\rho_2\bigr)^{u'-u'_1}\cr
\noalign{\bigskip}
\Bigl\{\circ-\bigl(\rho_2,-\rho_2\bigr)\Bigr\}\bigl(\rho_2\bigr)^r
&\circ&\bigl(-\rho_2\bigr)^{u'_1}\Bigl\{\circ-\bigl(-\rho_2,\rho_2\bigr)
\Bigr\}\bigl(-\rho_2\bigr)^{u'-u'_1}\cr
\noalign{\bigskip}
\bigl(-\rho_1\bigr)^{u-u_1}\Bigl\{\circ-\bigl(\rho_1,-\rho_1\bigr)\Bigr\}
\bigl(-\rho_1\bigr)^{u_1}&\circ&\bigl(\rho_1\bigr)^r\Bigl\{\circ-\bigl(
-\rho_1,\rho_1\bigr)\Bigr\}
\label{eq:subs1}
\eea
Here $\rho_1,\rho_2=\pm$, $r\geq1;\;u,u'\geq0$ and $0\leq u_1\leq u$,
$0\leq u'_1\leq u'$.
Remaining parts of subsequences larger than (\ref{eq:subs1}) are filled
by $c_1$ and/or $c_2$ with the choices $\ldots\rho_1$  and $\rho_2\ldots$
excluded for the left and right part in the case $u_1\neq u$ or $u'_1\neq
u'$, respectively. The third group lists all subsequences with two
single symbols $\circ$. They may be contained in two parts chosen
arbitrarily among
(\ref{eq:subs1}) or in one part $\circ(\rho_3)^{k_1}\bigl\{\circ-
(\rho,-\rho)\bigr\}(\rho_4)^{k_2}\circ$. The latter are obtained by
superimposing two parts (\ref{eq:subs1}) characterized by $\rho_i,u,u_1,
u',u'_1$ and $\tilde{\rho}_j,\tilde u,\tilde u_1,\tilde u',\tilde u'_1$
 such that the right factor
$\bigl\{\circ-(\rho_i,-\rho_i)\bigr\}$ of the left part coincides
with the left factor $\bigl\{\circ-(-\tilde{\rho}_j,\tilde{\rho}_j)
\bigr\}$ of the right part. Here the third expression in (\ref{eq:subs1})
or the first two expressions with $u'_1=u'$ may be taken for the 
left part and the second expression or the remaining two choices
with $\tilde u_1=\tilde u$ for the right part. For example,
the first option in (\ref{eq:subs1}) for both parts leads to
\bea
\bigl(-\rho_1\bigr)^{u-u_1}\Bigl\{\circ-\bigl(\rho_1,-\rho_1\bigr)
\Bigr\}\bigl(-\rho_1\bigr)^{u_1}\circ\bigl(-\rho_2\bigr)^{u'}\Bigl\{
\circ-\bigl(-\rho_2,\rho_2\bigr)\Bigr\}\bigl(\rho_2\bigr)^{\tilde u}
\circ\bigl(-\tilde{\rho}_2\bigr)^{\tilde u'_1}\Bigl\{\circ-\bigl(
-\tilde{\rho}_2,\tilde{\rho}_2\bigr)\Bigr\}\bigl(-\tilde{\rho}_2\bigr)^{
\tilde u'-\tilde u'_1}
\eea
with $u,u',\tilde u,\tilde u'\geq0$, $0\leq u_1\leq u$ and $0\leq
\tilde u'_1\leq\tilde u'$.
In any case, all parts $c_i$ occupying remaining space in the subsequence
are subject to the requirement specified below (\ref{eq:subs1}).
Alternatively, the two single $\circ$ may be found in a part
\bea
\bigl(-\rho\bigr)^{u-u_1}\Bigl\{\circ-\bigl(\rho,-\rho\bigr)\Bigr\}
\bigl(-\rho\bigr)^{u_1}\circ\bigl(\rho\bigr)^v\circ\bigl(-\rho\bigr)^{
u'_1}\Bigl\{\circ-\bigl(-\rho,\rho\bigr)\Bigr\}\bigl(-\rho\bigr)^{u'-u'_1}
\label{eq:subs2}
\eea
where $\rho=\pm$ and $v=0,1,2,\ldots$. If $u_1\neq u$ (or $u'_1\neq u'$),
any part $c_1,c_2$ left (right) of (\ref{eq:subs2}) must be different
from $\ldots\rho$ (or $\rho\ldots$).
Generally, the $n$-th group includes all subsequences with $n$ single
symbols $\circ$. The are distributed over the subsequence in parts
(\ref{eq:subs1}), (\ref{eq:subs2}) or in parts obtained by superimposing
two or more of these according to the above description. Here the part
(\ref{eq:subs2}) may be used with $u'_1=u'$ left 
of another part (\ref{eq:subs1}), (\ref{eq:subs2}) and with $u_1=u$ right of
another such part.
All parts $c_i$ completing the subsequence must be chosen within the
requirements stated below (\ref{eq:subs1}) and (\ref{eq:subs2}).
For $K=3,M=4$, $K=4,M=3$ or $M,K>3$, the subsequences collected this
way are not linearly independent. In particular, for suitably chosen
$a,b$, each term in the equation
\bea
\bigl(b\circ-a\bigr)-\bigl(b-\circ a\bigr)=\Bigl(b\Bigl\{\circ-\bigl(-+
\bigr)\Bigr\}-a\Bigr)-\Bigl(b-\Bigl\{\circ-\bigl(+-\bigr)\Bigr\}a\Bigr)
\label{eq:null1}
\eea
is found among the subsequences specified below equation (\ref{eq:start2}).
Such pairs of $a,b$ have the form
\bea
b=\ldots\Bigl\{\circ-\bigl(-+\bigr)\Bigr\}\qquad a=\Bigl\{\circ-\bigl(
+-\bigr)\Bigr\}\ldots
\label{eq:circ1a}
\eea
or
\bea
b=\ldots\Bigl\{\circ-\bigl(+-\bigr)\Bigr\}(-)^s\qquad a=(-)^t\Bigl\{
\circ-\bigl(-+\bigr)\Bigr\}\ldots
\label{eq:circ1b}
\eea
where $s,t=0,1,2,\ldots$. Similarly, each term in
\bea
\bigl(b+\circ-a\bigr)&=&\Bigl(b+\Bigl\{\circ-\bigl(-+\bigr)\Bigr\}-a
\Bigr)\cr
\noalign{\bigskip}
&&-\Bigl(b\Bigl\{\circ-\bigl(+-\bigr)\Bigr\}^2a\Bigr)+\Bigl(b\circ\Bigl\{
\circ-\bigl(+-\bigr)\Bigr\}a\Bigr)+\Bigl(b\Bigl\{\circ-\bigl(+-\bigr)
\Bigr\}\circ a\Bigr)-\bigl(b\circ\circ \;a\bigr)
\label{eq:null2}
\eea
is contained in the collection described above for appropriate pairs
$(a,b)$ with the form
\bea
b=\ldots\Bigl\{\circ-\bigl(+-\bigr)\Bigr\},\;a=(-)^s\Bigl\{\circ-\bigl(
-+\bigr)\Bigr\}\ldots\qquad\mbox{or}\qquad b=\ldots\Bigl\{\circ-\bigl(
-+\bigr)\Bigr\}(+)^t,\;a=\Bigl\{\circ-\bigl(+-\bigr)\Bigr\}\ldots
\label{eq:circ2}
\eea
with $s,t=0,1,2,\ldots$.
Finally, all terms of
\bea
&&\bigl(b\circ\circ-a\bigr)-\bigl(b-\circ\circ\;a\bigr)=
\Bigl(b\Bigl\{\circ-\bigl(-+\bigr)\Bigr\}^2-a\Bigr)-\Bigl(b-\Bigl\{\circ-
\bigl(+-\bigr)\Bigr\}^2a\Bigr)\cr
\noalign{\bigskip}
&&\qquad\qquad
-\Bigl(b\circ\Bigl\{\circ-\bigl(-+\bigr)\Bigr\}-a\Bigr)+\Bigl(b-\circ
\Bigl\{\circ-\bigl(+-\bigr)\Bigr\}a\Bigr)-\Bigl(b\Bigl\{\circ-\bigl(-+\bigr)
\Bigr\}\circ-a\Bigr)+\Bigl(b-\Bigl\{\circ-\bigl(+-\bigr)\Bigr\}\circ a\Bigr)
\label{eq:null3}
\eea
belong to the collection for suitable $a$ and $b$ satisfying
(\ref{eq:circ1b}). The lhs of equations (\ref{eq:null1}),
(\ref{eq:null2}) and (\ref{eq:null3}) can be removed from the groups of
subsequences introduced above.
The same applies to all subsequences obtained 
from (\ref{eq:null1})-(\ref{eq:null3}) by exchanging $+\leftrightarrow-$.

With respect to the resulting collection ${\cal K}'(K,M)$ of subsequences,
the elements $h^{(n,\kappa)}_{(a);\,(b)}$ 
form a triangular matrix. This is demonstrated by means
of a number $\Upsilon(a)\geq0$ introduced for such a subsequence $a$. 
The first contribution $\Upsilon_1(a)$ to this number 
counts all single symbols $\circ$ of $a$. Writing $a$ in the form
(\ref{eq:seq4}), this means $\Upsilon_1(a)=\sum_{l=0}^Rn_l$.
To define
$\Upsilon(a)$, each single $\circ$ in $a$ is substituted either by
$\bigl\{\circ-(-+)\bigr\}$ or $\bigl\{\circ-(+-)\bigr\}$. In each of these
$2^{\Upsilon_1(a)}$ subsequences $a'$, three types of
replacement are
carried out. The first type substitutes a part $-\rho\bigl\{\circ-(\rho,
-\rho)\bigr\}^L\rho$ found in $a'$
with $\rho=\pm,\;L=1,2,3,\ldots$ by $\bigl\{\circ 
-(\rho,-\rho)\bigr\}^{L+1}$. To such a replacement, the number $4L$ is
attributed. The second and third type substitutes $\rho\bigl\{\circ-
(-\rho,\rho)\bigr\}^L\bigl\{\circ-(\rho,-\rho)\bigr\}$ and
$\bigl\{\circ-(-\rho,\rho)\bigr\}\bigl\{\circ-(\rho,-\rho)\bigr\}^L\rho$
in $a'$
by $\bigl\{\circ-(-\rho,\rho)\bigr\}^L\rho\bigl\{\circ-(\rho,-\rho)\bigr\}$
and $\bigl\{\circ-(-\rho,\rho)\bigr\}\rho\bigl\{\circ-(\rho,-\rho)
\bigr\}^L$,
respectively. The number $2L$ is associated with each of these steps.
These three replacements are repeated until no part $\rho\bigl\{\circ-(-
\rho,\rho)\bigr\}$ or $\bigl\{\circ-(\rho,-\rho)\bigr\}\rho$ is left.
In general, starting from a given $a'$
several subsequences $\tilde a$ with this property can be
reached by different combinations of these steps. For any combination,
$\Upsilon_2^{a',\tilde a}(a)$ is introduced as the sum of the numbers
attributed to each replacement entailed in it. Furthermore, to any part
$\ldots\bigl\{\circ-(\rho,-\rho)\bigr\}(-\rho)^r\bigl\{\circ-(-\rho,\rho
)\bigr\}^{L'}(\rho)^s\bigl\{\circ-(\rho,-\rho)\bigr\}\ldots$ contained in
$\tilde a$ with $\rho=\pm,\;r,s=0,1,2,\ldots$ and $L'=1,2,3,\ldots$,
the number $(L'-1)^2$ is attributed. Finally,
denoting by $\Upsilon^{a',\tilde a}_3(a)$ the sum of all these numbers, 
$\Upsilon(a)$ is defined by
\bea
\Upsilon(a)\equiv\Upsilon_1(a)+\max_{a',\tilde a}\Bigl(
\Upsilon^{a',\tilde a}_2(a)+\Upsilon^{a',\tilde a}_3(a)\Bigr)
\label{eq:Ups}
\eea
In particular, this number takes the value zero for all subsequences
of the form (\ref{eq:eig1})-(\ref{eq:eig2}).
Equations (\ref{eq:jump1})-(\ref{eq:circre}) imply that
$\Upsilon(b)<\Upsilon(a)$ 
for any two subsequences $a,b$ of the collection
${\cal K}'(K,M)$ with $h^{(n,\kappa)}_{(a);\,(b)}\neq0$ and $a\neq b$. 
With respect to $h^{(n,\kappa)}$,
each number $h^{(n,\kappa)}_{(a);\,(a)}$ with $a\in{\cal K}'(K,M)$ is an
eigenvalue with the corresponding eigenvector given by a linear
combination of $a$ and some $b\in {\cal K}'(K,M)$ with
$h^{(n,\kappa)}_{(b);\,(b)}\neq h^{(n,\kappa)}_{(a);\,(a)}$ and
$h^{(n,\kappa)}_{(a);\,(b)}\neq0$ or $h^{(n,\kappa)}_{(a);\,(c_1)}h^{(n,
\kappa)}_{(c_1);\,(c_2)}\ldots h^{(n,\kappa)}_{(c_m);\,(b)}\neq0$ for
some $m>0$ and $c_m\in{\cal K}'(K,M)$. The coefficient of some
subsequence $b$ satisfying these properties may vanish as well.
An example is provided by the coefficient of $b=\bigl(\bigl\{\circ-(-+)
\bigr\}^2\bigr)$ in the linear combination starting with
$a=\bigl(+\bigl\{\circ-(-+)\bigr\}-\bigr)$.
The eigenvalue
$h^{(n,\kappa)}_{(a);\,(a)}$ depends on $n$ through the contribution
$-\kappa$. According to (\ref{eq:dia3}), the remaining dependence
on $n$ is given by $-\vartheta(a)\cdot n$, where $\vartheta(a)$
denotes the number of terms $\bigl\{\circ-(-+)\bigr\}$ and $\bigl\{
\circ-(+-)\bigr\}$ found in $a$ and is restricted by $1\leq
\vartheta(a)\leq \min(K,M)$.  
For given $M$ and $K$, the eigenvalues with the minimal or maximal
value of $\vartheta$ are readily classified.
All eigenvalues with $\vartheta(a)=1$ are given by $\kappa+n+r$ with
$1\leq r\leq\max(K,M)$ and by $\kappa+n+\max(K,M)+s$ with $1\leq s
\leq\min(K,M)$. Each eigenvalue $\kappa+n+r$ is found $2r-1$ times
for $1\leq r\leq \min(K,M)$. If $K\neq M$, the number of eigenvalues
$\kappa+n+r$ with $\min(K,M)<r\leq\max(K,M)$ is given by $2\min(K,M)$. The
eigenvalue $\kappa+n+\max(K;M)+s$ occurs $2\min(K,M)-2s+1$ times.
If $M=K$, the eigenvalues with $\vartheta(a)=M$ are $\kappa
+M(n+M)+r$, where $0\leq r\leq M$. These eigenvalues are
$\left({M\atop r}\right)$-fold degenerated. A similar pattern applies to
the case $\vartheta(a)=K<M$. For any set of numbers $\bigl\{r_l=0,1,2,
\ldots\bigr\}_{0\leq l\leq M}$
satisfying $\sum_{l=0}^Mr_l=M-K$ there is a set of
eigenvalues $\kappa+M(n+M)+r+\sum_{l=1}^Ml\,r _l$ with $0\leq r\leq M$.
Each of these eigenvalues is $\left({M\atop r}\right)$-fold degenerated.
Exchanging $K$ with $M$ yields the corresponding results in the case
$K>M$. 
The collection ${\cal K}'(K,M)$ can be made a basis for all subsequences
of the form (\ref{eq:seq3}) by adding the subsequence 
$\bigl((-)^{M-K}(\circ)^K\bigr)$ for $M\geq K$ and 
$\bigl((+)^{K-M}(\circ)^M\bigr)$ for $K\geq M$. According to
(\ref{eq:dia3})-(\ref{eq:circre}), the nonvanishing matrix elements
involving $\bigl((-)^{M-K}(\circ)^K\bigr)$ for $M\geq K$ read
\bea
h^{(n,\kappa)}_{\bigl((-)^{M-K}(\circ)^K\bigr);\;\bigl((-)^{M-K}(\circ)^K
\bigr)}&=&-\kappa\qquad M\geq K\cr
\noalign{\bigskip}
h^{(n,\kappa)}_{\bigl((+)^{K-M}(\circ)^M\bigr);\;\bigl((+)^{K-M}(\circ)^M
\bigr)}&=&-\kappa\qquad K\geq M
\label{eq;circonly1}
\eea
and
\bea
h^{(n,\kappa)}_{\bigl((-)^{M-K}(\circ)^K\bigr);\;\bigl((-)^{M-K}(\circ
)^{K-1-L}\bigl\{\circ-(\rho,-\rho)\bigr\}(\circ)^L\bigr)}&=&-(n+1+
\delta_{\rho,-}+2L)\qquad M\geq K\cr
\noalign{\bigskip}
h^{(n,\kappa)}_{\bigl((+)^{K-M}(\circ)^M\bigr);\;\bigl((+)^{K-M}(\circ
)^{M-1-L}\bigl\{\circ-(\rho,-\rho)\bigr\}(\circ)^L\bigr)}&=&-(n+1+
\delta_{\rho,-}+2L)\qquad K\geq M
\label{eq:circonly2}
\eea
for $\rho=\pm$ and $0\leq L\leq \min(K,M)-1$. As shown in appendix A,
all subsequences $\bigl((-)^{M-K}(\circ)^{K-1-L}
\bigl\{\circ-(\rho,-\rho)\bigr\}(\circ)^L\bigr)$
and $\bigl((+)^{K-M}(\circ)^{M-1-L}
\bigl\{\circ-(\rho,-\rho)\bigr\}(\circ)^L\bigr)$
can be written as linear combinations of subsequences contained in the
collection ${\cal K}'(K,M)$. Obviously, $h^{(n,\kappa)}$ is trigonal
with respect to the enlarged collection
${\cal K}(K,M)\equiv\bigl\{{\cal K}'(K,M),\bigl((-)^{M-K}(\circ)^K\bigr)
\bigr\}$ for $M\geq K$ or 
${\cal K}(K,M)\equiv\bigl\{{\cal K}'(K,M),\bigl((+)^{K-M}(\circ)^M\bigr)
\bigr\}$ for $K\geq M$. An eigenvector of $H_0$ with the eigenvalue
$h^{(n,\kappa)}_{(a);\,(a)}$ is given by a linear combination of the
particular configuration containing the subsequence $a$ and the
infinitely many
configurations obtained from it by repeated action of $H_0$ with
the diagonal element of $H_0$ different from $h^{(n,\kappa)}_{(a);\,
(a)}$.

\subsection{General configurations}
\label{sec:genset}

Now the structure of a half-infinite configuration $\bigl(\ldots,i_3
j^*_3,i_2j^*_2,i_1j^*_1\bigr)$ with several sets of subsequences composed
of $1^*1\pm0^*0$ and/or $0^*1,1^*0$ can be specified. 
A general
configuration may be written as linear combination of configurations
\bea
&&\Bigl(\bigl\{a_l(K_l,M_l)\bigr\}_{1\leq l\leq T},\;\bigl\{j_{s+1},i_s
\bigr\}_{s>m_T},\;\bigl\{j_{s+1},i_s\bigr\}_{m_l<s\leq m_{l+1},\,1\leq
l<T},\;\bigl\{j_{s+1},i_s\bigr\}_{1\leq s\leq m_1},\;j_1\Bigr)\equiv\cr
\noalign{\bigskip}
&&\Bigl(
\ldots i_{m_T+2},j^*_{m_T+2}i_{m_T+1},a_T(K_T,M_T),j^*_{m_T+1}i_{m_T},
j^*_{m_T}\ldots i_{m_3+2},j^*_{m_3+2}i_{m_3+1},a_3(K_3,M_3),j^*_{m_3+1}
i_{m_3},j^*_{m_3}\ldots\cr
\noalign{\bigskip}
&&\ldots
i_{m_2+2},j^*_{m_2+2}i_{m_2+1},a_2(K_2,M_2),j^*_{m_2+1}i_{m_2},j^*_{m_2}
\ldots i_{m_1+2},j^*_{m_1+2}i_{m_1+1},a_1(K_1,M_1),j^*_{m_1+1}i_{m_1},
j^*_{m_1}\ldots ,j^*_2i_1,j^*_1\Bigr)
\label{eq:confgen}
\eea
where $M_l,K_l\geq1$, $m_{l+1}>m_l$ and $a_l(K_L,M_L)$ denotes
any subsequence chosen from the basis ${\cal K}(K_l,M_l)$. The parts
$\ldots i_{m_T+2},j^*_{m_T+2}i_{m_T+1}$ and $j^*_{m_{l+1}+1}i_{m_{l+1}},
j^*_{m_{l+1}}
\ldots i_{m_l+2},j^*_{m_l+2}, i_{m_l+1}$ with $1\leq l\leq T-1$ as well as
$j^*_{m_1+1}i_{m_1},j^*_{m_1}\ldots i_2,j^*_2i_1,j^*_1$ do not contain
any of the terms (\ref{eq:ndsubs}). Only finitely many entries differ
from $2$. In analogy to the case $T=1$, a number 
\bea
\Upsilon
\Bigl(\bigl\{a_l(K_l,M_l)\bigr\}_{1\leq l\leq T},\;\bigl\{j_{s+1},i_s
\bigr\}_{s>m_T},\;\bigl\{j_{s+1},i_s\bigr\}_{m_l<s\leq m_{l+1},\,1\leq
l<T},\;\bigl\{j_{s+1},i_s\bigr\}_{1\leq s\leq m_1},\;j_1\Bigr)\equiv
\sum_{l=1}^T\Upsilon\bigl(a_l(K_l,M_l)\bigr)
\label{eq:Upgen}
\eea
with $\Upsilon\bigl(a_l(K_l,M_l)\bigr)$ defined by (\ref{eq:Ups})
may be introduced for the configuration (\ref{eq:confgen}).

Two configurations
$A=
\Bigl(\bigl\{a'_l(K'_l,M'_l)\bigr\}_{1\leq l\leq T'},\;\bigl\{j'_{s+1},i'_s
\bigr\}_{s>m_T'},\;\bigl\{j'_{s+1},i'_s\bigr\}_{m'_l<s\leq m'_{l+1},\,1\leq
l<T'},\;\bigl\{j'_{s+1},i'_s\bigr\}_{1\leq s\leq m'_1},\;j'_1\Bigr)$ and
$A'=
\Bigl(\bigl\{a_l(K_l,M_l)\bigr\}_{1\leq l\leq T},\;\bigl\{j_{s+1},i_s
\bigr\}_{s>m_T},\;\bigl\{j_{s+1},i_s\bigr\}_{m_l<s\leq m_{l+1},\,1\leq
l<T},\;\bigl\{j_{s+1},i_s\bigr\}_{1\leq s\leq m_1},\;j_1\Bigr)$
related by $h_{A;\,A'}\neq0$ can differ in five ways. First,
the second configuration may be obtained from the first by replacing
one subsequence $a_l(K_l,M_l)\in{\cal K}(K_l,M_l)$ by another subsequence
$a'_l(K_l,M_l)\in{\cal K}(K_l,M_l)$. Then
\bea
\Upsilon(A')<\Upsilon(A)
\label {eq:ord1}
\eea
Second, the two sets of subsequences $\bigl\{a_l(K_l,M_l)\bigr\}_{l\leq T}$
and $\bigl\{a'_l(K'_l,M'_l)\bigr\}_{l\leq T'}$ coincide but some entries
$i_s,j_s$ differ. Third, an additional subsequence $a(K,M)\in{\cal K}
(K,M)$ together with a suitable change in the entries $i_s,j_s$ may
occur in the second configuration while all subsequences $a_l(K_l,
M_l)$ of the first configuration are kept unchanged.
Alternatively, one subsequence $a_l(K_l,M_l)$ of the first configuration
is substituted by additional entries $i_s,j_s$ in the second configuration
while all remaining subsequences of the first configuration are left
unchanged. The additional entries $i_s,j_s$ don't give rise
to any terms (\ref{eq:ndsubs}).
All other cases involve a substitution of one subsequence $a_l(K_l,M_l)
\in{\cal K}(K_l,M_l)$ (or of two neighboring subsequences $a_{l+1}(K_{l+1}
,M_{l+1})\in{\cal K}(K_{l+1},M_{l+1})$
and $a_l(K_l,M_l)\in{\cal K}(K_l,M_l)$)
by a subsequence from a different collection
${\cal K}(K'_l,M'_l)$ (or by subsequences from different collections
${\cal K}(K'_{l+1},M'_{l+1})$ and ${\cal K}(K'_l,M'_l)$). 
This may be accompanied by adjustments in some entries $i_s,j_s$ not
producing any terms (\ref{eq:ndsubs}).
In all cases except the first one, 
\bea
\Omega(A')>\Omega(A)
\label{eq:ord2}
\eea
with $\Omega(B)$ defined by (\ref{eq:ome}).
Hence, the elements $h_{A;\,A'}$ with $A$, $A'$ of type (\ref{eq:confgen}),
(\ref{eq:seq3}) or without any sequence (\ref{eq:ndsubs})
form a triangular matrix if arranged in an order indicated by 
(\ref{eq:ord1}) and (\ref{eq:ord2}).

To each decomposition $a_l(K_L,M_L)=\bigl(b'\bigl\{
\circ-(\rho,-\rho)\bigr\}b\bigr)\in{\cal K}(K,M)$
the number $m_l+1+\delta_{\rho,-}
+\lambda(b)+\sum_{l'=1}^{l-1}(K_{l'}+M_{l'})$
with $\rho=\pm$ and $\lambda(b)$ as defined below (\ref{eq:seq4})
is attributed. The sum of these numbers for all such decompositions of
$a_l(K_L,M_L)$ may be denoted by $\gamma(a_l)$. Then the diagonal element
of $H_0$ with respect to the configuration (\ref{eq:confgen}) is given by
\bea
&-&\sum_{l=1}^T\gamma(a_l)-\sum_{l=2}^T
\sum_{l'=1}^{l-1}(K_{l'}+M_{l'})\,\bigl(
\delta_{j_{m_l+1},2}+\delta_{i_{m_l+1},2}+\delta_{i_{m_l},0}+\delta_{
j_{m_l+2},0}\bigr)\cr
\noalign{\bigskip}
&-&\sum_{l=1}^T\Bigl((m_l+1)\delta_{j_{m_l+1},2}
+\bigl(m_l+K_l+M_l)\bigr)\,\delta_{i_{m_l+1},2}
+m_l\,\delta_{i_{m_l},0}+\bigl(m_{l}+1+K_l+M_l)\bigr)
\,\delta_{j_{m_l+2},0}\Bigr)\cr
\noalign{\bigskip}
&-&\sum_{t=1}^{m_1}t\bigl(
(1-\delta_{t,m_1})y_{i_{t+1},j_{t+1},i_t}+y_{j_t,i_t,j_{t+1}}
\bigr)\cr
\noalign{\bigskip}
&-&\sum_{l=1}^{T-1}\sum_{t=m_l+1}^{m_{l+1}}\Bigl(t+\sum_{k=1}^l(K_k+M_k
)\Bigr)
\bigl((1-\delta_{t,m_{l+1}})y_{i_{t+1},j_{t+1},
i_t}+(1-\delta_{t,m_{l}+1})y_{j_t,i_t,j_{t+1}}\bigr)\cr
\noalign{\bigskip}
&-&\sum_{t=m_T+1}^{\infty}\Bigl(t+\sum_{k=1}^T(K_k+M_k)\Bigr)\bigl(
y_{i_{t+1},j_{t+1},i_t}+(1-\delta_{t,m_T+1})y_{j_t,i_t,j_{t+1}}
\bigr)
\label{eq:diagen}
\eea
If $T=1$, the second terms of the first and third line in
(\ref{eq:diagen}) are dropped. A configuration without terms
(\ref{eq:ndsubs}) has its diagonal element given by (\ref{eq:tri}) and
(\ref{eq:beta}).
Each diagonal element of $H_0$ on a configuration $A$ written in the form
(\ref{eq:confgen}) is an eigenvalue
of $H_0$. The corresponding eigenvector is a linear combination of $A$
and the configurations with a different diagonal element arising from
repeated action of $H_0$ on $A$.

The values of the diagonal elements (\ref{eq:beta}) and (\ref{eq:diagen})
have upper bounds $0$ and $-2$, respectively. Three vanishing diagonal
elements 
are found. As stated above, the corresponding configurations are
$(\ldots,2^*2,2^*2,i)$ with $i=0,1,2$. For a fixed value of $h_{A;\,A}$,
only finitely many configurations $A$ exist. Their structure will
be investigated in the following section.

\section{The module $V(\Lambda_2)$}
\label{sec:mod}

Choosing a value $N$ for a given configuration $(\ldots,j_3^*i_2,j_2^*
i_1,j_1^*)$ such that $j_n=2\forall n>N+1$ and $i_n=2\forall n>N$, the
numbers 
\bea
\bar h_1\bigl((\ldots,j^*_3i_2,j^*_2i_1,j^*_1)\bigr)&=&-\delta_{j_{N+1},0}
-\delta_{j_{N+1},1}+\sum_{n=1}^N\bigl(\delta_{i_n,0}+\delta_{i_n,1}-
\delta_{j_n,0}-\delta_{j_n,1}\bigr)\cr
\noalign{\bigskip}
\bar h_2\bigl((\ldots,j^*_3i_2,j^*_2i_1,j^*_1)\bigr)&=&\delta_{j_{N+1},1}
+\delta_{j_{N+1},2}+\sum_{n=1}^N\bigl(\delta_{j_n,1}+\delta_{j_n,2}-
\delta_{i_n,1}-\delta_{i_n,2}\bigr)
\label{eq:hdef}
\eea
may be used to define an action of $h_1$ and $h_2$ on
$(\ldots,j^*_3i_2,j^*_2i_1,j^*_1)$.
In the following, for each configuration the numbers $\bar h_1$,
$\bar h_2$ and value $\bar H_0$ of the diagonal
element of $H_0$ on  $(\ldots,j^*_3i_2,j^*_2i_1,j^*_1)$ will be collected
writing $(\bar H_0,\bar h_1,\bar h_2)$. For the configurations $(\ldots,
2^*2,2^*2,i^*)$ the definitions (\ref{eq:hdef}) yield 
$(0,0,1)$, $(0,-1,1)$ and $(0,-1,0)$.
According to the remarks at the end of the last section, all other
configurations have lower values $\bar H_0$.  From (\ref{eq:beta}),
the configurations with the value $\bar H_0=-1$ are $\bigl(\ldots,2^*2,
2^*2,2^*i,j^*)$, $(\ldots,2^*2,2^*2,1^*i,j^*)$, $(\ldots,2^*2,2^*2,1^*2,j^*
)$ and $(\ldots,2^*2,2^*2,0^*0,j^*)$ with $i,j=0,1$. To these,
(\ref{eq:hdef}) assigns the values $(-1,-2,0)$, $(-1,-2,1)$ and
$(-1,0,k)$, $(-1,-1,k)$ with $k=0,1,2$, where $(-1,0,1)$ and
$(-1,-1,0)$ are twofold and $(-1,-1,1)$ is threefold degenerated.

These values may be compared to the weight components associated to
weight components of the irreducible module $V(\Lambda_2)$ at level one. 
The latter is characterized by a unique highest weight state $\vert
\kappa\rangle$ with the properties
\bea
e_i\,\vert\kappa\rangle=0\qquad\qquad h_i\,\vert\kappa\rangle=\delta_{i,2}
\,\vert\kappa\rangle\qquad \mbox{for}\;i=0,1,2
\label{eq:hwdef}
\eea
All states in the module are generated by the action of $f_i$ on
$\vert\kappa\rangle$. The eigenvalues of $h_i$ on such a state may be
denoted by $\lambda_i$, $i=0,1,2$. They provide
the coefficients of $\Lambda_i$ in the expansion of an
affine weight in terms of the fundamental weights and $\delta$.
At level one, the coefficient
$\lambda_0$ of $\Lambda_0$ is given by $1-\lambda_1-\lambda_2$. To
specify a weight, the three coefficients are written in the form
$[\lambda_0,\lambda_1,\lambda_2]$. With the action of the grading
operator $d$ on the highest weight state fixed by $d\,\vert\kappa\rangle
=0$, the eigenvalues of d on any weight state are called its grade.
According to (\ref{eq:hwdef}) and the defining relations
(\ref{eq:def1})-(\ref{eq:ddef}),
the following weights are found in $V(\Lambda_2)$ at grade $0$ and
$-1$:
\bea
\begin{array}{ccccccccccc}
&&&&&&&[-1,0,2]_1&&\cr
\noalign{\smallskip}
&&&&&&\swarrow&&\searrow&\cr
\noalign{\smallskip}
[0,0,1]_1&&&&&[0,0,1]_2&&&&[0,-1,2]_1\cr
\noalign{\smallskip}
&\searrow&&&\swarrow&&\searrow&&\swarrow&\cr
\noalign{\medskip}
&&[1,-1,1]_1\qquad\qquad\qquad\qquad&[1,0,0]_1&&&&[1,-1,1]_3&&\cr
\noalign{\smallskip}
&\swarrow&&&\searrow&&\swarrow&&\searrow&\cr
\noalign{\smallskip}
[2,-1,0]_1&&&&&[2,-1,0]_2&&&&[2,-2,1]_1\cr
\noalign{\smallskip}
&&&&&&\searrow&&\swarrow&\cr
\noalign{\smallskip}
&&&&&&&[3,-2,0]_1&&
\end{array}
\eea
Here the left and right part of the diagram refers to the weight
space at grade $0$ and $-1$, respectively, and the subscripts denote
the multiplicity of the weights. Arrows pointing southwest (southeast)
indicate the action of $f_1$ ($f_2$). 
Identifying the grade of a weight state
with the value $\bar H_0$ associated to the configurations listed above,
a one-to-one correspondence between the pairs $\lambda_1,\lambda_2$ and
the values $\bar h_1,\bar h_2$ is found at grade $0$ and $-1$.
This correspondence applies to the next three lower grades, too.
 
The states of $V(\Lambda_2)$ at a fixed nonvanishing grade can be
arranged arranged in a finite number of separate sets labeled by
pairs $(\lambda_1,\lambda_2)$ with
$\lambda_1\geq0$ and $\lambda_2>0$. A set labeled by will be called
$\pi(\lambda_1,\lambda_2)$ in the following. It contains $4(\lambda_1+
\lambda_2)$ states with weights$(1-\lambda_1-\lambda_2,\lambda_1,
\lambda_2)$, $(1+2\mu-\lambda_1-\lambda_2,\lambda_1-
\mu,\lambda_2-\mu)$, $(2\mu-\lambda_1-\lambda_2,\lambda_1-\mu+1,\lambda_2
-\mu)$ and $(2\mu-\lambda_1,-\lambda_2,\lambda_1-\mu,\lambda_2-\mu+1)$
with $0<\mu\leq\lambda_1+\lambda_2$. The multiplicity is two for each
weight $(1+2\mu-\lambda_1-\lambda_2,\lambda_1-\mu,\lambda_2-\mu)$ with
$0<\mu<\lambda_1+\lambda_2$ and one for all others. For the four lowest
nonvanishing grades, the pairs $(\lambda_1,\lambda_2)$ are listed in
the following table with the multiplicity specified by a subscript:
\bea
\begin{array}{llll}
\mbox{grade}-1\qquad\qquad&\mbox{grade}\,-2\qquad\qquad
&\mbox{grade}\,-3\qquad\qquad&\mbox{grade}\,-4\cr
\noalign{\bigskip}
\noalign{\smallskip}
\;[0,2]_1&\;[1,2]_1&\;[1,2]_2&\;[1,3]_1\cr
\noalign{\medskip}
\;[0,1]_1&\;[0,2]_2&\;[0,3]_1&\;[1,2]_5\cr
\noalign{\medskip}
&\;[0,1]_2&\;[1,1]_1&\;[0,3]_2\cr
\noalign{\medskip}
&\;[-1,2]_1&\;[0,2]_5&\;[1,1]_3\cr
\noalign{\medskip}
&&\;[0,1]_4&\;[0,2]_{10}\cr
\noalign{\medskip}
&&\;[-1,2]_2&\;[0,1]_8\cr
\noalign{\medskip}
&&&\;[-1,2]_5\cr
\noalign{\medskip}
&&&\;[-1,3]_1
\end{array}
\eea
Collecting all configurations with $\bar H_0$ given by $-2,-3,-4$
reveals that the number of configurations with values $(\bar H_0,
\bar h_1,\bar h_2)$ coincides with the multiplicity of the weight
$(1-\bar h_1-\bar h_2,\bar h_1,\bar h_2)$ at grade $\bar H_0$. Thus
the one-to-one correspondence between CTM-configurations and
weight states of the module $V(\lambda_2)$ at level one is confirmed
down to grade $-4$.

Equation (\ref{eq:diagen}) is involved in the evaluation for the diagonal
element of $H_0$ on the configurations $(\ldots,2^*2,2^*2,2\circ0^*)$ and
$(\ldots,2^*2,2^*2,2\circ1^*)$ at grade $-2$, on 
$(\ldots,2^*2,2^*2,2\circ2^*)$, $(\ldots,2^*2,2^*2\bigl\{\circ-(+-)\bigr\}
j^*)$, $(\ldots,2^*2,2^*2,2^*i\circ j^*)$, $(\ldots,2^*2,2^*1,0\circ j^*)$,
$(\ldots,2^*2,2^*2,+\circ j^*)$, $(\ldots,2^*2,2^*2,-\circ j^*)$ and
$(\ldots,2^*2,2^*2\circ1^*,2j^*)$ with $i,j=0,1$ at grade $-3$ and on
$57$ configurations at grade $-4$.

Due to (\ref{eq:beta}), a configuration $(\ldots,2^*2,
2^*2,j^*_{n+1}i_n,j^*_ni_{n-1},\ldots,j^*_3i_2,j^*_2i_1j^*_1)$
with $j^*_{n+1}\neq2$ or $j^*_{n+1}=2, i_n\neq2$
not containing any subsequence (\ref{eq:ndsubs}) has the diagonal element
of $H_0$ bound by $-n(n+1)\leq \bar H_0$.
In particular, the value of
the lower bound is taken for the configuration $\bigl(\ldots,2^*2,2^*2,2^*
(0,2^*)^n\bigr)$. As is easily verified from (\ref{eq:beta}) and
(\ref{eq:diagen}), all other configurations with the values $(\bar H_0,
n,n+1)$ satisfy $\bar H_0<-n(n+1)$. The corresponding state in $V(\Lambda_2
)$ is
$E^{2,+}_{-n}\ldots E^{2,+}_{-2}E^{2,+}_{-1}\,E^{1,+}_{-n}\ldots
E^{1,+}_{-2}E^{1,+}_{-1}\,\vert\kappa\rangle$ with weight
$(-2n,n,n+1)$ and grade $-n(n+1)$. At level one, $f_0\,\vert\kappa\rangle
=0$. Because of this property and (\ref{eq:def5}),
any other state in $V(\Lambda_2)$ with the same weight has a lower grade.

Generally, the weights $(-m-n,m,n+1)$ and $(2+m+n,
-n-1,-m)$ with $n\geq m\geq0$
appear in the module $V(\Lambda_2)$ at grades bounded from above
by $-{1\over2}n(n+1)-{1\over2}m(m+1)$. At grade
$\leq-{1\over2}n(n+1)-{1\over2}m(m+1)$ the multiplicity is one.
In terms of Drinfeld generators, the corresponding states read
\bea
E^{2,+}_{-m}\ldots E^{2,+}_{-2}E^{2,+}_{-1}\;E^{1,+}_{-n}\ldots
E^{1,+}_{-2}E^{1,+}_{-1}\,\vert\kappa\rangle\qquad\mbox{and}\qquad
(f_1f_2)^{m+n+1}\,
E^{2,+}_{-m}\ldots E^{2,+}_{-2}E^{2,+}_{-1}\;E^{1,+}_{-n}\ldots
E^{1,+}_{-2}E^{1,+}_{-1}\,\vert\kappa\rangle
\label{eq:set1}
\eea
Besides, the weights $(1-n,-1,n+1)$ and $(1+n,-n-1,1)$ are found at
grade $-{1\over2}n(n+1)$ with multiplicity one.
The corresponding states $E^{2,-}_0\,E^{1,+}_{-n}
\ldots E^{1,+}_{-2}E^{1,+}_{-1}\,\vert\kappa\rangle$ and $E^{2,-}_0
(E^{1,-}_0E^{2,-}_0)^n\,E^{1,+}_{-n}\ldots E^{1,+}_{-2}E^{1,+}_{-1}\,
\vert\kappa\rangle$ belong to the set $\pi(0,n+1)$ at this grade.
For the grade $-{1\over2}n(n+1)-1$, the multiplicities of the weights
$(-n-1,1,n+1)$, $(-n,0,n+1)$, $(1-n,-1,n+1)$ and $(2-n,-2,n+1)$ are
$1$, $3$, $3$ and $1$, respectively. Hence, the last two weights are
contained in a set $\pi(-1,n+1)$ with the weight
$(1-n,-1,n+1)$ related to the state $E^{2,-}_{-1}\,E^{1,+}_{-n}\ldots
E^{1,+}_{-2}E^{1,+}_{-1}\,\vert\kappa\rangle$. Similarly, sets 
$\pi(-m,n+1)$ with $n\geq m>0$ are present at any grade $\leq -{1\over2}n
(n+1)-{1\over2}m(m+1)$. Exactly one set is found at grade $-{1\over2}n
(n+1)-{1\over2}m(m+1)$. The state corresponding to its weights are
generated by acting with $f_1$ and $f_2$ on the states
\bea
E^{2,-}_{-m}\ldots E^{2,-}_{-2}E^{2,-}_{-1}\;E^{1,+}_{-n}\ldots E^{1,+}_{
-2}E^{1,+}_{-1}\,\vert\kappa\rangle\qquad n\geq m>0
\label{eq:set2}
\eea
For $n\geq m\geq0$, the weights $(1+2k+m-n,-m-k-1,n+1-k)$ with $0\leq
k\leq n+m$ contained in the sets labeled by $(-m,n+1)$ have multiplicity
one at the maximal grade.

In addition, sets $\pi(n+m+1,n+1)$ with $m,n\geq0$ appear
at all grades $\leq-(n+1)(n+m+1)-{1\over2}(m+1)(m+2)-\delta_{n,0}$.
Again, there  is exactly one set at the maximal grade.
The states related to its weights arise from the
action of $f_1$ and $f_2$ on
\bea
\begin{array}{ll}
E^{1,-}_{-1}\;E^{2,+}_{-1}E^{1,+}_{-1}\,\vert\kappa\rangle&n=m=0\cr
\noalign{\bigskip}
E^{1,-}_{-1}\;E^{2,+}_{-(m+2)}\ldots E^{2,+}_{-4}E^{2,+}_{-3}\;E^{2,+}_{
-1}E^{1,+}_{-1}\,\vert\kappa\rangle&n=0,\;m>0\cr
\noalign{\bigskip}
E^{2,+}_{-(n+m+2)}\ldots E^{2,+}_{-(n+3)}E^{2,+}_{-(n+2)}\;E^{2,+}_{-n}
\ldots E^{2,+}_{-2}E^{2,+}_{-1}\;E^{1,+}_{-n}\ldots E^{1,+}_{-2}E^{1,+}_{
-1}\,\vert\kappa\rangle\qquad\qquad&n>0,\;m\geq0
\end{array}	
\label{set3}
\eea
At the maximal grade $-(n+1)(n+m+1)-{1\over2}(m+1)(m+2)-\delta_{n,0}$,
the weights $(n+m-k,n-k)$ with $n>0,\;m\geq0$ and $0\leq k\leq 2n+m$
found in the sets $(n+m,n+1)$ have multiplicity one. The complete
collection of all states with multiplicity one present in
the level-one module $V(\Lambda_2
)$ is obtained by adding the zero-grade states $\vert\kappa\rangle$ and
$E^{1,-}_0E^{2,-}_0\,\vert\kappa\rangle$ to those listed so fare.

The weights belonging to the sets $\pi(m,n)$ for $m\geq0$ and
$n>0$ exhaust all weights in the module $V(\Lambda_2)$ at level one.
The upper bounds for the grades with a set $\pi(m,n)$ with
$n>m\geq0$
present should be compared with the maximal diagonal element of $H_0$
for the configurations with eigenvalues $\bar h_1=m$, $\bar h_2=n$
and $\bar h_1=-n$, $\bar h_2=-m$.
From (\ref{eq:beta}) and (\ref{eq:diagen}),
these are the configurations 
\bea
\bigl(\ldots,2^*2,2^*2,2^*(0,1^*)^{n-m}(0,2^*)^m\bigr)\qquad\mbox{and}
\qquad
\bigl(\ldots,2^*2,2^*2,2^*(2,1^*)^{n-m}(2,0^*)^{m+1}\bigr)\qquad n>m\geq0
\label{eq:con1}
\eea
with $\bar H_0=-{1\over2}n(n+1)-{1\over2}m(m+1)$. There are no other
configurations with the same values $(\bar H_0,\bar h_1,\bar h_2)$.
Therefore the configurations (\ref{eq:con1}) can be viewed as
counterparts of the states (\ref{eq:set1}).

Moreover, configurations with $\bar h_1=-m-1-k$, $\bar h_2=n+1-k$
with $0\leq k\leq n+m$ are found
only with $\bar H_0\leq-{1\over2}n(n+1)-{1\over2}m(m+1)$. The maximal
value of $\bar H_0$ is attributed to the configurations
\bea
\bigl(\ldots,2^*2,2^*2,1^*(0,1^*2,1^*)^m(2,1^*)^k(0,1^*)^{n-m-k}\bigr)
\qquad\qquad
n\geq m\geq 0,\;0\leq k\leq n+m
\label{eq:con2}
\eea
Thus the configurations (\ref{eq:con2}) can be related to
the states $E^{2,-}_0(E^{1,-}_0E^{2,-}_0)^k\;E^{1,+}_{-n}\ldots E^{1,+}_{-2}
E^{1,+}_{-1}\,\vert\kappa\rangle$ for $m=0$ and to
$E^{2,-}_0\,(E^{1,-}_0E^{2,-}_0)^kE^{2,-}_{-m}\ldots
E^{2,-}_{-2}E^{2,-}_{-1}\;
E^{1,+}_{-n}\ldots E^{1,+}_{-2}E^{1,+}_{-1}\,\vert\kappa\rangle
$ for $m>0$. The weights of these states
are part of the sets $\pi(-m,n+1)$ at grade
$-{1\over2}n(n+1)-{1\over2}m(m+1)$ with $n\geq m\geq0$.

Configurations with $\bar h_1=m+1-k$ and $\bar h_2=-k$ with $m\geq0$
and $0\leq k\leq m+1$ exist with $\bar H_0\leq -{1\over2}(m+2)(m+3)$
with the maximal value of $\bar H_0$ valid for the configurations
\bea
\bigl(\ldots,2^*2,2^*2,2^*(1,0^*)^{k+1}(1,2^*)^{m+1-k}\bigr)\qquad m\geq0,
\;0\leq k\leq m+1
\label{eq:con3}
\eea
Finally, configurations with $\bar h_1=n+m-k$ and $\bar h_2=n-k$ with
$n>0, m\geq0$ and $0\leq k\leq m+2n$ are present only with
$\bar H_0\leq -(n+m)(n+1)- {1\over2}m(m+1)$.
In the case $n=1$, the maximal value of $\bar H_0$ is attributed to the
configurations
\bea
\begin{array}{ll}
\bigl(\ldots,2^*2,2^*2,2^*(1,2^*)^{m+1}\bigr)\qquad\qquad&\mbox{for}\;
k=0\cr
\noalign{\bigskip}
\bigl(\ldots,2^*2,2^*2,0^*(1,0^*)^{k-1}(1,2^*)^{m+2-k}\bigr)
\qquad\qquad&\mbox{for}\;1\leq k\leq m+1\cr
\noalign{\bigskip}
\bigl(\ldots,2^*2,2^*2,2^*(1,0^*)^{m+1}\bigr)&\mbox{for}\;k=m+2
\end{array}
\label{eq:con4}
\eea
with $m\geq0$. If $n>1$, the maximal value of $\bar H_0$ applies to the 
configurations
\bea
\begin{array}{ll}
\bigl(\ldots,2^*2,2^*2,2^*(1,2^*)^{m+1}(0,2^*)^{n-1}\bigr)&\mbox{for}\;
k=0\cr
\noalign{\bigskip}
\bigl(\ldots,2^*2,2^*2,0^*(1,0^*)^{k-1}(1,2^*)^{m+2-k}(0,2^*)^{n-1}
\bigr)\qquad\qquad&\mbox{for}\;1\leq k\leq m+1\cr
\noalign{\bigskip}
\bigl(\ldots,2^*2,2^*2,0^*(1,0^*)^{m}(2,0^*)^l2,2^*1,2^*(0,2^*)^{n
-l-2}\bigr)\qquad\qquad&\mbox{for}\;k=m+2l+2,\;0\leq l\leq n-2\cr
\noalign{\bigskip}
\bigl(\ldots,2^*2,2^*2,0^*(1,0^*)^{m}(2,0^*)^{l+1}1,2^*(0,2^*)^{n
-l-2}\bigr)\qquad\qquad&\mbox{for}\;k=m+2l+3,\;0\leq l\leq n-2\cr
\noalign{\bigskip}
\bigl(\ldots,2^*2,2^*2,0^*(1,0^*)^{m}(2,0^*)^{n-1}1,0^*\bigr)&\mbox{for}
\;k=m+2n
\end{array}
\label{eq:con5}
\eea
with $m\geq0$. The configurations (\ref{eq:con3}), (\ref{eq:con4})
and (\ref{eq:con5})
are counterparts of the states with weights $(1+2k-2n-m,n+m-k,n-k)$
in the set labeled by $(m+n,n+1)$
at grade $-(n+m)(n+1)-{1\over2}m(m+1)$ for any $n,m\geq0$, $n+m>0$ .

As is easily verified from (\ref{eq:beta})  and (\ref{eq:diagen}), the
values $(\bar H_0,\bar h_1,\bar h_2)$ for a configuration with $\bar H_0$
determined by (\ref{eq:diagen}) are shared by at least two different
configurations. If (\ref{eq:beta}) covers the evaluation of $\bar H_0$
for a configuration containing a subsequence $(1^*1+0^*0)$, the value of
$\bar H_0$ does not change when replacing this subsequence by $(1^*1-0^*0)$.
The values $\bar h_1,\,\bar h_2$ for any other configuration with
$\bar H_0$ determined by (\ref{eq:beta}) can be attributed to a weight
in one of the sets $\pi(m,n)$
related to the configurations (\ref{eq:con1})-(\ref{eq:con5}) as
specified above. Again the grade coincides with the value $\bar H_0$.
For example, $\bigl(\ldots,2^*2,2^*(2,0^*)^k2,1^*(0,2^*)^{n-k}\bigr)$
with $0\leq k\leq n$ and $\bigl(\ldots,2^*2,2^*(2,0^*)^k2,1^*2,2^*(0,2^*)^{
n-k-2}\bigr)$ with $0\leq k\leq n-1$ correspond to the weights
$(1+4k-2n,n-2k-1,n-2k+1)$ and $(3+4k-2n,n-2k-2,n-2k)$ of the set
$\pi(n,n+1)$ at grade $-n(n+1)$, $n>1$.	
Due to the weight structure of the sets, any values $(\bar H_0,\bar h_1,
\bar h_2)$ different from $\bigl(-{1\over2}n(n+1)-{1\over2}m(m+1),m,n+1
\bigr)$, $\bigl(-{1\over2}n(n+1)-{1\over2}m(m+1),-n-1,-m\bigr)$ with
$n\geq m\geq0$ or $\bigl(-{1\over2}n(n+1)-{1\over2}m(m+1),-m-k-1,n+1-k
\bigr)$ with $n\geq m\geq0$, $0\leq k\leq n+m$ or $\bigl(-(n+m)(n+1)
-{1\over2}m(m+1)-\delta_{n,0},n+m-k,n-k\bigr)$ with $n,m\geq0$, $n+m>0$
and $0\leq k\leq 2n+m$ or $(0,-k,1-k)$ with $k=0,1$ occur more than once.

Thus each state contained in the irreducible level-one module
$V(\Lambda_2)$ with multiplicity can be mapped onto exactly one of
the CTM-configurations with  a nondegenerated triple $(\bar H_0,
\bar h_1, \bar h_2)$ and vice versa.
This observation as well as the result on all states at grade $0$ to
$-4$ support the conjecture that the CTM-configurations $(\ldots,j^*_3
i_2,j^*_2i_1,j^*_1)$ and the weight states of $V(\Lambda_2)$ at level
one are in one-to-one correspondence.

\section{Appendix A}
\label{sec:app}

All subsequences generated by the action
of $h^{(n,\kappa)}$ on $\bigl((-)^{
M-K}(\circ)^K\bigr)$ for $M\geq K$ or from $\bigl((+)^{K-M}(\circ)^M\bigr)$
for $K\geq M$ have to be expressible
as linear combinations of the same terms and subsequences contained
in the groups specified by (\ref{eq:start1})-(\ref{eq:null3}),
if ${\cal K}(K,M)=\bigl\{{\cal K}'(K,M),\bigl((-)^{M-K}(\circ)^K\bigr)
\bigr\}$ for
$M\geq K$ or ${\cal K}(K,M)=\bigl\{{\cal K}'(K,M),\bigl((+)^{K-M}(\circ)^M
\bigr)\bigr\}$
for $K\geq M$ is to provide a basis of all subsequences of the form
(\ref{eq:seq3}) with the same values of $K$ and $M$.

It is convenient to consider the case $K=M$ first. The relevant
subsequences are then given by $\bigl((\circ)^{M-1-L}\bigl\{\circ-
(\rho,-\rho)\bigr\}(\circ)^L\bigr)$ with $\rho=\pm$ and $0\leq L\leq M-1$.
For $K=M=3$, the required rewritings read
\bea
\bigl(\bigl\{\circ-(+-)\bigr\}\circ\circ\bigr)&=&-\bigl(\bigl\{\circ-(+-)
\bigr\}\bigl\{\circ-(-+)\bigr\}^2\bigr)+\bigl(\bigl\{\circ-(+-)\bigr\}
\circ\bigl\{\circ-(-+)\bigr\}\bigr)+\bigl(\bigl\{\circ-(+-)\bigr\}\bigl\{
\circ-(-+)\bigr\}\circ\bigr)\cr
\noalign{\bigskip}
&&-\bigl(\bigl\{\circ-(+-)\bigr\}-\bigl\{\circ-(+-)\bigr\}+\bigr)+\bigl(
\bigl\{\circ-(+-)\bigr\}-\circ+\bigr)
\label{eq:ex1}
\eea
and
\bea
\bigl(\circ\circ\bigl\{\circ-(+-)\bigr\}\bigr)&=&-\bigl(\bigl\{\circ-(-+)
\bigr\}^2\bigl\{\circ-(+-)\bigr\}\bigr)+\bigl(\bigl\{\circ-(-+)\bigr\}
\circ\bigl\{\circ-(+-)\bigr\}\bigr)+\bigl(\circ\bigl\{\circ-(-+)\bigr\}
\bigl\{\circ-(+-)\bigr\}\cr
\noalign{\bigskip}
&&-\bigl(-\bigl\{\circ-(+-)\bigr\}+\bigl\{\circ-(+-)\bigr\}\bigr)+\bigl(
-\circ+\bigl\{\circ-(+-)\bigr\}\bigr)
\label{eq:ex2}
\eea
Exchanging $+\leftrightarrow-$ in (\ref{eq:ex1}), (\ref{eq:ex2}) yields
$\bigl(\bigl\{\circ-(-+)\bigr\}\circ\circ\bigr)$ and $\bigl(\circ\circ
\bigl\{\circ-(-+)\bigr\}\bigr)$ in terms of subsequences of
${\cal K}'(3,3)$. These expressions and (\ref{eq:ex1}), (\ref{eq:ex2})
may be combined to rewrite a subsequence built from several terms
$(\circ)^k$ with $k=1,2$ separated by $\bigl\{\circ-(-+)\bigr\}$ and/or
$\bigl\{\circ-(+-)\bigr\}$ in terms of subsequences contained in
${\cal K}'(M,M)$. The equations
\bea
\bigl(a\bigl\{\circ-(-+)\bigr\}(+)^s\circ(-)^r\bigr)&=&\bigl(a\bigl\{\circ
-(-+)\bigr\}(+)^{s+1}\circ(-)^{r-1}\bigr)+\bigl(a\bigl\{\circ-(-+)\bigr\}
(+)^s\bigl\{\circ-(+-)\bigr\}(+)^r\bigr)\cr
\noalign{\bigskip}
&&-\bigl(a\bigl\{\circ-(-+)\bigr\}(+)^{s+1}\bigl\{\circ-(-+)\bigr\}(+)^{r
-1}\bigr)
\label{eq:ex3}
\eea
with $s\geq0$, $r\geq1$ and a part $a$ as described below (\ref{eq:seq4})
allow to reformulate $\bigl(\bigl\{\circ-(+-)\bigr\}\circ\circ\,(+)^r\bigr)$
in terms of subsequences found in ${\cal K}'(r+3,3)$:
\bea
&&\bigl(\bigl\{\circ-(+-)\bigr\}\circ\circ\,(+)^r\bigr)=-\bigl(\bigl\{\circ-
(+-)\bigr\}\bigl\{\circ-(-+)\bigr\}^2\bigr)+\bigl(\bigl\{\circ-(+-)\bigr\}
\circ\bigl\{\circ-(+-)\bigr\}(+)^r\bigr)\cr
\noalign{\bigskip}
&&\qquad\qquad+\bigl(\bigl\{\circ-(+-)\bigr\}
\bigl\{\circ-(-+)\bigr\}(+)^r\circ\bigr)
+\sum_{t=0}^{r-1}\bigl(\bigl\{\circ-(+-)\bigr\}\bigl\{\circ-(-+)\bigr\}
(+)^t\bigl\{\circ-(+-)\bigr\}(+)^{r-t}\bigr)\cr
\noalign{\bigskip}
&&\qquad\qquad-\sum_{t=1}^r\bigl(\bigl\{
\circ-(+-)\bigr\}\bigl\{\circ-(-+)\bigr\}(+)^t\bigl\{\circ-(-+)\bigr\}
(+)^{r-t}\bigr)\cr
\noalign{\bigskip}
&&\qquad\qquad-\bigl(\bigl\{\circ-(+-)\bigr\}-\bigl\{\circ-(+-)\bigr\}(
+)^{r+1}\bigr)+\bigl(\bigl\{\circ-(+-)\bigr\}-\circ(+)^{r+1}\bigr)
\label{eq:ex4a}
\eea
Furthermore, $\bigl\{\circ-(+-)\bigr\}$ in (\ref{eq:ex1}) and
the leftmost term $\bigl\{\circ-(+-)\bigr\}$ in (\ref{eq:ex4a})
can be substituted by $\bigl\{\circ-(+-)\bigr\}(-)^s$ with $s=1,2,3,\ldots$.
Reversing the order of symbols and exchanging $+\leftrightarrow-$ provides
expressions for $\bigl((-)^r\circ\circ\,(+)^s\bigl\{\circ-(+-)\bigr\}\bigr)$,
$\bigl(\bigl\{\circ-(-+)\bigr\}(+)^s\circ\circ\,(-)^r\bigr)$ and $\bigl((+)^r
\circ\circ\,(-)^s\bigl\{\circ-(-+)\bigr\}\bigr)$.
Equation (\ref{eq:ex1}) may be generalized starting from
\bea
&&\Bigl(\bigl\{\circ-(+-)\bigr\}(\circ)^{M-1}\Bigr)+(-1)^n\Bigl(\bigl\{
\circ-(+-)\bigr\}(-+)^{M-1}\Bigr)=\cr
\noalign{\bigskip}
&&\qquad\Bigl(\bigl\{\circ-(+-)\bigr\}\bigl\{\circ-(-+)\bigr\}^{M-1}\Bigr)+
\cr
\noalign{\bigskip}
&&\qquad+\sum_{r=1}^{M-2}(-1)^r\sum_{k_1=0}^{M-r-1}\qquad\sum_{k_2=0}^{
M-r-1-k_1}
\qquad
\sum_{k_3=0}^{M-r-1-k_1-k_2}\ldots\sum_{k_r=0}^{M-r-1-k_1-\ldots-k_{r-1}}\cr
\noalign{\bigskip}
&&
\Bigl(\bigl\{\circ-(+-)\bigr\}\bigl\{\circ-(-+)\bigr\}^{k_1}\circ
\bigl\{\circ-(-+)\bigr\}^{k_2}\circ\ldots
\circ\bigl\{\circ-(-+)\bigr\}^{k_r}\circ\bigl\{\circ-(-+)\bigr\}^{M-r-1-k_1-
\ldots-k_r}\Bigr)
\label{eq:ex4}
\eea
Again, the leftmost term $\bigl\{\circ-(+-)\bigr\}$ may be replaced by
$\bigl\{\circ-(+-)\bigr\}-$.
In order to obtain an expression for $\bigl(\bigl\{\circ-(+-)\bigr\}
(\circ)^{M-1}+\bigr)$ or $\bigl(\bigl\{\circ-(+-)\bigr\}-(\circ)^{M-1}+
\bigr)$ by inserting a symbol $+$ left of the right border of (\ref{eq:ex4}),
an appropriate
rewriting of $\bigl(a\bigl\{\circ-(-+)\bigr\}(\circ)^n+\bigr)$ is
required:
\bea
&&\Bigl(a\bigl\{\circ-(-+)\bigr\}(\circ)^n+\Bigr)+(-1)^n\Bigl(a
\bigl\{\circ-(-+)\bigr\}(+-)^n+\Bigr)=\cr
\noalign{\bigskip}
&&\qquad=\Bigl(a\bigl\{\circ-(-+)\bigr\}\bigl\{\circ-(+-)\bigr\}^n+
\Bigr)+\cr
\noalign{\bigskip}
&&\qquad+\sum_{r=1}^{n-1}(-1)^r\sum_{k_1=0}^{n-r}
\qquad\sum_{k_2=0}^{n-r-k_1}\qquad
\sum_{k_3=0}^{n-r-k_1-k_2}\ldots\sum_{k_r=0}^{n-r-k_1-k_2-\ldots-k_{r-1}}
\cr
\noalign{\bigskip}
&&\Bigl(a\bigl\{\circ-(-+)\bigr\}\bigl\{\circ-(+-)
\bigr\}^{k_1}\circ\bigl\{\circ-(+-)\bigr\}^{k_2}\circ
\ldots\circ\bigl\{\circ-(+-)\bigr\}^{k_{r}}\circ\bigl\{\circ
-(+-)\bigr\}^{n-r-k_1-\ldots-k_{r}}+\Bigr)
\label{eq:ex5}
\eea
On the other hand,
\bea
&&\Bigl(a\bigl\{\circ-(-+)\bigr\}+(\circ)^n\Bigr)+(-1)^n\Bigl(a
\bigl\{\circ-(-+)\bigr\}+(-+)^n\Bigr)=\cr
\noalign{\bigskip}
&&\qquad=\Bigl(a\bigl\{\circ-(-+)\bigr\}+\bigl\{\circ-(-+)\bigr\}^n
\Bigr)+\cr
\noalign{\bigskip}
&&\qquad+\sum_{r=1}^{n-1}(-1)^r\sum_{k_1=0}^{n-r}
\qquad\sum_{k_2=0}^{n-r-k_1}\qquad
\sum_{k_3=0}^{n-r-k_1-k_2}\ldots\sum_{k_r=0}^{n-r-k_1-k_2-\ldots-k_{r-1}}
\cr
\noalign{\bigskip}
&&\Bigl(a\bigl\{\circ-(-+)\bigr\}+\bigl\{\circ-(-+)
\bigr\}^{k_1}\circ\bigl\{\circ-(-+)\bigr\}^{k_2}\circ
\ldots\circ\bigl\{\circ-(-+)\bigr\}^{k_{r}}\circ\bigl\{\circ
-(-+)\bigr\}^{n-r-k_1-\ldots-k_{r}}\Bigr)
\label{eq:ex6}
\eea
A reformulation for the first term on the lhs of (\ref{eq:ex6})
is provided by replacing the left most term $\bigl\{\circ-(+-)\bigr\}$
in (\ref{eq:ex4}) by  $\bigl\{\circ-(+-)\bigr\}-$, switching all
signs and adding the piece $a$ at the left border.
Combining this with equations (\ref{eq:ex5}) and (\ref{eq:ex6})
leads to an expression of $\bigl(\bigl\{\circ-(+-)\bigr\}-(\circ)^{M-2}
+\bigr)$ analogous to the equation (\ref{eq:ex4}) for $\bigl(\bigl\{
\circ-(+-)\bigr\}(\circ)^{M-1}\bigr)$. Use of this expression in
\bea 
&&\Bigl(\bigl\{\circ-(+-)\bigr\}-(\circ)^{M-2}+\Bigr)-(-1)^n\Bigl(\bigl\{
\circ-(+-)\bigr\}(-+)^{M-1}\Bigr)=\cr
\noalign{\bigskip}
&&\qquad\Bigl(\bigl\{\circ-(+-)\bigr\}-\bigl\{\circ-(+-)\bigr\}^{M-2}+
\Bigr)\cr
\noalign{\bigskip}
&&\qquad-\sum_{r=1}^{M-3}(-)^r\sum_{k_1=0}^{M-r-2}
\qquad\sum_{k_2=0}^{M-r-2-
k_1}\qquad\sum_{k_3=0}^{M-r-2-k_1-k_2}\ldots\sum_{k_r=0}^{M-r-2-k_1-\ldots-
k_{r-1}}\cr
\noalign{\bigskip}
&&\Bigl(\bigl\{\circ-(+-)\bigr\}-\bigl\{\circ-(+-)\bigr\}^{k_1}\circ\bigl\{
\circ-(+-)\bigr\}^{k_2}\circ\ldots\circ\bigl\{\circ-(+-)\bigr\}^{k_{r}}
\bigl\{\circ-(+-)\bigr\}^{M-r-2-k_1-\ldots-k_{r}}\Bigr)
\label{eq:ex7}
\eea
and comparison with (\ref{eq:ex4}) yields an rewriting of $\bigl(
\bigl\{\circ-(+-)\bigr\}(\circ)^{M-1}\bigr)$ in terms of subsequences
involving lower powers of $\circ$. Analogous expressions for
$\bigl((\circ)^{M-1}\bigl\{\circ-(+-)\bigr\}\bigr)$, $\bigl(\bigl\{
\circ-(-+)\bigr\}(\circ)^{M-1}\bigr)$ and 
$\bigl((\circ)^{M-1}\bigl\{\circ-(-+)\bigr\}\bigr)$ are obtained by
reversing the order of symbols and/or exchanging $+\leftrightarrow-$.
These may be combined to provide reformulations of subsequences
containing powers $(\circ)^m$ with $m\leq M-1$ separated by $\bigl\{\circ
-(-+)\bigr\}$ and/or $\bigl\{\circ-(+-)\bigr\}$. Together with
(\ref{eq:ex1})-(\ref{eq:ex4a}), this ensures the existence of rewritings of
$\bigl(a\bigl\{\circ-(\rho,-\rho)\bigr\}(\circ)^n\bigr)$ and 
$\bigl((\circ)^n\bigl\{\circ-(\rho,-\rho)\bigr\}a\bigr)$ in terms
of subsequences of ${\cal K}'(K,M)$ with $K,M$ according to the
piece $a$. Hence all subsequences $\bigl((\circ)^{M-1-L}\bigl\{\circ-
(\rho,-\rho)\bigr\}(\circ)^L\bigr)$ can be reformulated as linear
combinations of subsequences found in ${\cal K}(M,M)$. Continuing
along this lines it is straightforward to give such rewritings for any
subsequence of the form (\ref{eq:seq3}) with $K=M$.

For $M>K$, all subsequences $\bigl((-)^{M-K-t_0-t_1-\ldots-t_K}\circ
(-)^{t_K}\circ\ldots(-)^{t_2}\circ(-)^{t_1}\circ(-)^{t_0}\bigr)$
with $t_l\geq0\;\forall l$ and $M-K-\sum_{l=0}^Kt_l\geq0$ can be
rewritten in terms of $\bigl((-)^{M-K}(\circ)^K\bigr)$ and subsequences
of ${\cal K}'(K,M)$. This is easily demonstrated in the case $K=1$:
\bea
&&
\bigl((-)^{M-1-L}\circ(-)^L\bigr)-\bigl((-)^{M-2-L}\circ(-)^{L+1}\bigr)=\cr
\noalign{\bigskip}
&&\qquad\qquad\qquad\qquad\qquad\qquad=
\bigl((-)^{M-1-L}\bigl\{\circ-(+-)\bigr\}(-)^L\bigr)-\bigl((-)^{M-2-L}
\bigl\{\circ-(-+)\bigr\}(-)^{L+1}\bigr)
\label{eq:ex8}
\eea
for $0\leq L\leq M-2$. Hence all subsequences can be obtained by adding
$\bigl((-)^{M-1}\circ\bigr)$ to the collection ${\cal K}'(1,M)$. Of
course, any other $\bigl((-)^{M-1-L}\circ(-)^L\bigr)$ would be appropriate
as well for supplementing ${\cal K}'(1,M)$.
Similarly, for $K=2$,
\bea
&&\bigl((-)^{M-3-L}\circ(-)^{L+1}\circ\bigr)-\bigl((-)^{M-2-L}\circ(-)^L
\circ\bigr)=\cr
\noalign{\bigskip}
&&\qquad\qquad\qquad\qquad
=\bigl((-)^{M-3-L}\bigl\{\circ-(-+)\bigr\}(-)^{L+1}\circ\bigr)-\bigl(
(-)^{M-2-L}\bigl\{\circ-(+-)\bigr\}(-)^L\circ\bigr)
\label{eq:ex9}
\eea
for $0\leq L\leq M-3$. The rhs of (\ref{eq:ex9}) can be expressed in terms
of subsequences of ${\cal K}'(2,M)$ using
\bea
&&\bigl((-)^{M-2-L}\bigl\{\circ-(-+)\bigr\}(-)^{L-L'}\circ(-)^{L'}\bigr)=
\bigl((-)^{M-2-L}\bigl\{\circ-(-+)\bigr\}(-)^{L-L'-1}\circ(-)^{L'+1}\bigr)
+\cr
\noalign{\bigskip}
&&+\bigl((-)^{M-2-L}\bigl\{\circ-(-+)\bigr\}(-)^{L-L'}\bigl\{\circ-(+-)
\bigr\}(-)^{L'}\bigr)-\bigl((-)^{M-2-L}\bigl\{\circ-(-+)\bigr\}(-)^{L-L'+1}
\bigl\{\circ-(-+)\bigr\}(-)^{L'+1}\bigr)
\label{eq:ex10}
\eea
with $0\leq L\leq M-2$ and $0\leq L'\leq L$. The order of symbols may be
reversed. Thus, any $\bigl((-)^{M-2-L}\circ(-)^L\circ\bigr)$ or
$\bigl(\circ(-)^L\circ(-)^{M-2-L}\bigr)$ can be expressed by
$\bigl((-)^{M-2}\circ\circ\bigr)$ and ${\cal K}'(2,M)$. Moreover, the
rhs of 
\bea
&&\bigl((-)^{M-3-L-L'}\circ(-)^{L+1}\circ(-)^{L'}\bigr)-\bigl((-)^{M-2-
L-L'}\circ(-)^L\circ(-)^{L'}\bigr)=\cr
\noalign{\bigskip}
&&\qquad\qquad\qquad=\bigl((-)^{M-3-L-L'}\bigl\{\circ-(-+)\bigr\}(-)^{L
+1}\circ(-)^L\bigr)-\bigl((-)^{M-2-L-L'}\bigl\{\circ-(+-)\bigr\}(-)^L
\circ(-)^{L'}\bigr)
\label{eq:ex11}
\eea 
with $L'\geq1$ and $0\leq L\leq M-3-L'$ can be rewritten in terms of
the same set of subsequences by repeated use of
\bea
&&\bigl((-)^{M-2-L}\bigl\{\circ-(+-)\bigr\}(-)^{L-L'}\circ(-)^{L'}\bigr)=
\bigl((-)^{M-2-L}\bigl\{\circ-(+-)\bigr\}(-)^{L-L'+1}\circ(-)^{L'-1}\bigr)
+\cr
\noalign{\bigskip}
&&+\bigl((-)^{M-2-L}\bigl\{\circ-(+-)\bigr\}(-)^{L-L'}\bigl\{\circ-(-+)
\bigr\}(-)^{L'}\bigr)-\bigl((-)^{M-2-L}\bigl\{\circ-(+-)\bigr\}(-)^{L-
L'+1}\bigl\{\circ-(+-)\bigr\}(-)^{L'-1}\bigr)
\label{eq:ex12}
\eea
for $0\leq L\leq M-2$ and $1\leq L'\leq L$. The order of symbols may be
reversed in (\ref{eq:ex11}) and (\ref{eq:ex12}). This
accounts for the required rewriting of
any subsequence $\bigl((-)^{M-2-L-L'}\circ(-)^L\circ(-)^{L'}\bigr)$
with $0\leq L\leq M-2$ and $0\leq L'\leq M-2-L$. 
Continuing with this procedure the statement is readily extended to
all subsequences of the form (\ref{eq:seq3}) with $K=2$. Upon
repeated application of
\bea
\bigl(b-\circ\,a\bigr)-\bigl(b\circ\,-a\bigr)=\bigl(b-\bigl\{\circ-(+-)
\bigr\}a\bigr)-\bigl(b\bigl\{\circ-(-+)\bigr\}-a\bigr)
\eea
the above procedure for $K=2$ generalizes in a straightforward manner
to general $K<M$.
Switching all signs and exchanging $M\leftrightarrow K$ yields the
corresponding formulae for the case $K>M$.

\end{document}